\documentclass[aps, pra, reprint, twocolumn,superscriptaddress]{revtex4-2}
\usepackage{amsmath,amssymb, bm}
\usepackage{amsthm}
\usepackage{amssymb}
\usepackage{booktabs}
\usepackage{graphicx}
\usepackage{braket} 
\usepackage{xcolor}
\usepackage{tikz}
\usepackage{tikz-3dplot}
\usetikzlibrary{3d}
\usepackage{float}
\usepackage{microtype}
\usepackage{ragged2e}
\newcounter{myfigurecount}
\newcommand{\mycaption}[1]{
	\stepcounter{myfigurecount} 
	\par\medskip
	\noindent
	\small
	\textbf{FIG. \themyfigurecount.} #1 
	\par\medskip
}

\usepackage[colorlinks=true,
citecolor=blue,
linkcolor=blue,
urlcolor=black]{hyperref}

\begin{document}
	
	\sloppy
	
	\title{Construction of Sets of Orthogonal Quantum States with Minimal Nonlocality in Bipartite and Tripartite Systems of Unequal Local Dimensions}
	\author{Ben-Hao \surname{Zhao}}%
    \affiliation{College of Computer Science and Engineering, Shandong University of Science and Technology, Qingdao, 266590, China}

   \author{Dong-Huan \surname{Jiang}}
   \affiliation{College of Mathematics and Systems Science, Shandong University of Science and Technology, Qingdao, 266590, China}
   
   \author{Yu-Guang \surname{Yang}}%
   \affiliation{College of Cyberspace Science and Technology, Beijing University of Technology, Beijing, 100124, China}

   \author{Guang-Bao  \surname{Xu}}%
   \email{xu\_guangbao@163.com}
   \affiliation{College of Computer Science and Engineering, Shandong University of Science and Technology, Qingdao, 266590, China}

	\date{\today}
	
	\begin{abstract}
  The research on minimal nonlocality aims to determine the minimal cardinality of nonlocal sets of quantum states. However, the construction of a nonlocal set of states in bipartite or tripartite quantum systems with unequal local dimensions remains unsolved. In this paper, we first give a method to construct a set of orthogonal quantum states with minimal nonlocality in $\mathbb{C}^{4} \otimes \mathbb{C}^{7}$  quantum system. Then we give a general method to construct a set of orthogonal quantum states with minimal nonlocality in a bipartite quantum system with unequal local dimensions. Furthermore, we generalize the construction method to tripartite quantum system with unequal local dimensions, and construct a set of orthogonal quantum states with minimal nonlocality in  $\mathbb C^{d_1} \otimes \mathbb C^{d_2}\otimes \mathbb C^{d_3}$ quantum system for $5\le d_1 < d_2 < d_3$. Our work settles the construction problem of a set of orthogonal states with minimal nonlocality in both bipartite and tripartite systems with unequal local dimensions.
	\end{abstract}
	
	\maketitle
	
	\section{Introduction}
	The local distinguishability of quantum states is a fundamental research topic in quantum information theory. In practical applications such as quantum communication and distributed computing, a set of orthogonal quantum states can be perfectly distinguished via global measurements \cite{1,2}. It is particularly critical to adopt local operations and classical communication (LOCC) to reduce the overhead of state transmission in quantum communication.
	
	Initially, people intuitively believed that a set of orthogonal product states (OPSs) could be distinguished. However, in 1999, Bennett et al. \cite{3} first discovered a set of orthogonal product states, which cannot be perfectly distinguished by LOCC. This phenomenon is known as quantum nonlocality without entanglement (QNWE). Since then, many scholars have carried out in-depth research on this phenomenon. Feng and Shi \cite{4} fully characterized the state sets with QNWE in $\mathbb C^3\otimes \mathbb C^3$ and $\mathbb C^2\otimes \mathbb C^2\otimes \mathbb C^2$ systems. Bhattacharya et al. \cite{5} extended QNWE to the framework of generalized probabilistic theories. The above research has stimulated researchers' interest in the study of QNWE.
	
	As research advances, many results on local distinguishability of quantum state sets have been put forward. Walgate et al. \cite{6} proved that two orthogonal pure states can be exactly distinguished by LOCC. Chen et al. \cite{7} analyzed the connection between orthogonality and local distinguishability for arbitrary state sets, derived a universal necessary condition for LOCC distinguishability of a class of orthogonal states. Over the past decade, constructing sets of orthogonal quantum states that are locally indistinguishable has become a key research focus. The research scope has mostly expanded from bipartite systems to multipartite systems, and numerous innovative construction methods have been proposed continuously \cite{8,9,10,11,12,13,14,15,16,17,18,19,20,21,22,23,24,25}. In recent years, with the deepening of research on nonlocal sets of orthogonal quantum states, many scholars have further advanced the study of indistinguishability from various perspectives. Halder et al. \cite{26} put forward the concept of strong nonlocality. Constructing OPS sets and orthogonal entangled state sets with strong quantum nonlocality has attracted extensive research interest \cite{27,28,29,30,31,32}.
	
	To better depict the nonlocality of OPS sets, in 2023, Zhu et al. \cite{33} first proposed the concept of minimal nonlocality. Minimal nonlocality corresponds to a specific nonlocal orthogonal quantum state set. A set satisfies the condition of minimal nonlocality when there exists a state within the set whose elimination renders all remaining states perfectly distinguishable under LOCC. Zhu et al. illustrated that this concept can characterize the lower bound of the number of states contained in nonlocal OPS sets. Meanwhile, they put forward a  method to construct completable minimal nonlocal orthogonal product state sets. They constructed minimal nonlocal sets with \(3d-1\) and \(3d\) states in \(\mathbb{C}^{d} \otimes \mathbb{C}^{d}\) quantum system, and extended this construction approach to tripartite and multipartite quantum systems. The proposal of this concept is of great significance in quantum information theory. It verifies that LOCC can accurately distinguish different quantum states under specific scenarios, provides core support for the encoding and decoding of quantum communication and quantum information processing, and facilitates the development of efficient communication protocols. 
	
	Although research on minimal nonlocality has achieved certain progress, prominent limitations still exist. Existing construction approaches are only applicable to equal-dimensional bipartite or multipartite systems \cite{34}, and a method for constructing minimal nonlocality in unequal-dimensional systems has not yet been proposed. In this paper, we first construct a set of quantum states with minimal nonlocality in  $\mathbb{C}^{4} \otimes \mathbb{C}^{7}$  quantum system. And then we give a general method to construct a set of orthogonal quantum states with minimal nonlocality in $\mathbb C^{d_1} \otimes \mathbb C^{d_2}$ quantum system  for $4\le d_1 < d_2$. Next, we propose a method to construct  a set of orthogonal quantum states with minimal nonlocality in  $\mathbb C^{d_1} \otimes \mathbb C^{d_2}\otimes \mathbb C^{d_3}$ quantum system for $5\le d_1 < d_2 < d_3$. The rest of this paper is organized as follows. In Sec.~\ref{sec:2}, some necessary preliminaries are given. In Sec.~\ref{sec:3}, we elaborate the construction method of sets with minimal nonlocality in bipartite unequal-dimensional systems and give a proof for minimal nonlocality. In Sec.~\ref{sec:4}, we extend the proposed method to tripartite unequal-dimensional systems and prove the minimal nonlocality of the resulting states. Finally, in Sec.~\ref{sec:5}, a brief conclusion is given.

	\section{PRELIMINARIES}
	\label{sec:2}
	In this section, we provide some definitions, which will be used in what follows. Note that the notation $\frac{1}{\sqrt{n}} |i_1 \pm i_2 \pm \dots \pm i_n\rangle$ denotes $\frac{1}{\sqrt{n}} (|i_1\rangle \pm |i_2\rangle \pm \dots \pm |i_n\rangle)$ in this paper. For the convenience of subsequent proofs, the quantum states in this work are not normalized.
	
	\textit{Definition 1.} \cite{35} If a set of orthogonal quantum states cannot be distinguished by LOCC, then we say it has the property of local indistinguishability or we say the set is locally indistinguishable or nonlocal.
	
	\textit{Definition 2.} \cite{36} A measurement is considered trivial if it does not obtain any useful information for identifying the quantum states. Conversely, a measurement is considered nontrivial if it provides valuable information that can be used to discern or characterize the quantum states.
	
	\textit{Definition 3.} \cite{33} A nonlocal set $\Omega$ of quantum states is said to have minimal nonlocality if there exists a state $|\phi\rangle \in \Omega$ such that $\Omega - \{|\phi\rangle\}$ can be perfectly distinguished by LOCC.
	
	\textit{Lemma 1.}(Kramer's rule \cite{37}) \label{lemma1} A system of equations
	\[
	\begin{cases}
		\alpha_{11}x_1 + \alpha_{12}x_2 + \cdots + \alpha_{1n}x_n = \beta_1\\
		\alpha_{21}x_1 + \alpha_{22}x_2 + \cdots + \alpha_{2n}x_n = \beta_2\\
		\quad\vdots\\
		\alpha_{n1}x_1 + \alpha_{n2}x_2 + \cdots + \alpha_{nn}x_n = \beta_n
	\end{cases}
	\]
	has a unique solution if its coefficient determinant
	\\
	\\
	\[
	\begin{vmatrix}
		\alpha_{11} & \alpha_{12} & \cdots & \alpha_{1n}\\
		\alpha_{21} & \alpha_{22} & \cdots & \alpha_{2n}\\
		\vdots & \vdots & \ddots & \vdots\\
		\alpha_{n1} & \alpha_{n2} & \cdots & \alpha_{nn}
	\end{vmatrix}
	\neq 0,
	\]
	where $\alpha_{\lambda\mu}$ and $\beta_\lambda$ are complex numbers for $\lambda = 1,2,\dots,n$ and $\mu = 1,2,\dots,n$.

	\section{Minimal Nonlocality of Orthogonal Quantum State Sets in Bipartite Unequal-Dimensional Quantum Systems}
	\label{sec:3}
	In this section, we mainly investigate the construction of orthogonal entangled state sets with minimal nonlocality in $\mathbb{C}^{d_1} \otimes \mathbb{C}^{d_2}$ system ($4\leq d_1 < d_2$). For intuitive comprehension, we first construct a nonlocal orthogonal quantum state set in $\mathbb{C}^{4} \otimes \mathbb{C}^{7}$ quantum system and verify its minimal nonlocality. On this basis, we further generalize the construction to the general $\mathbb{C}^{d_1} \otimes \mathbb{C}^{d_2}$ system with $4\leq d_1 < d_2$, and similarly prove that the constructed set also possesses minimal nonlocality.
	
	\begin{widetext}
		\centering
		\begin{tikzpicture}[
			box/.style = {
				draw,
				line width=0.8mm,
				minimum size=1.4cm,
				font=\large\bfseries
			},
			c1/.style = {fill=red!25},
			c2/.style = {fill=blue!15},
			c3/.style = {fill=green!15},
			c4/.style = {fill=orange!15},
			c5/.style = {fill=purple!25},
			c6/.style = {fill=cyan!15},
			c7/.style = {fill=pink!15},
			c8/.style = {fill=yellow!25},
			c9/.style = {fill=gray!25}
			]
			
			\node[box, c9] at (0, 0)    {9};
			\node[box, c4] at (1.4, 0)  {4};
			\node[box, c5] at (2.8, 0)  {5};
			\node[box, c3] at (4.2, 0)  {3};
			\node[box, c6] at (5.6, 0)  {6};
			\node[box, c7] at (7.0, 0)  {7};
			\node[box, c8] at (8.4, 0)  {8};
			
			\node[box, c1] at (0, -1.4) {1};
			\node[box]      at (1.4, -1.4) {};
			\node[box]      at (2.8, -1.4) {};
			\node[box, c4] at (4.2, -1.4) {4};
			\node[box, c6] at (5.6, -1.4) {6};
			\node[box, c7] at (7.0, -1.4) {7};
			\node[box, c8] at (8.4, -1.4) {8};
			
			\node[box, c2] at (0, -2.8) {2};
			\node[box]      at (1.4, -2.8) {};
			\node[box]      at (2.8, -2.8) {};
			\node[box, c5] at (4.2, -2.8) {5};
			\node[box]      at (5.6, -2.8) {};
			\node[box]      at (7.0, -2.8) {};
			\node[box]      at (8.4, -2.8) {};
			
			\node[box, c3] at (0, -4.2) {3};
			\node[box, c1] at (1.4, -4.2) {1};
			\node[box, c2] at (2.8, -4.2) {2};
			\node[box]      at (4.2, -4.2) {};
			\node[box]      at (5.6, -4.2) {};
			\node[box]      at (7.0, -4.2) {};
			\node[box, c9] at (8.4, -4.2) {9};
			
			\node at (0, 1)   {$|0\rangle$};
			\node at (1.4, 1) {$|1\rangle$};
			\node at (2.8, 1) {$|2\rangle$};
			\node at (4.2, 1) {$|3\rangle$};
			\node at (5.6, 1) {$|4\rangle$};
			\node at (7.0, 1) {$|5\rangle$};
			\node at (8.4, 1) {$|6\rangle$};
			
			\node at (-1, 0)    {$|0\rangle$};
			\node at (-1, -1.4) {$|1\rangle$};
			\node at (-1, -2.8) {$|2\rangle$};
			\node at (-1, -4.2) {$|3\rangle$};
			
			\node[font=\bfseries] at (4.2, 1.5) {Bob};
			\node[font=\bfseries, rotate=90] at (-1.7, -2.1) {Alice};
		\end{tikzpicture}
		
		\mycaption{The structure of the set $S_1$ in $ \mathbb C^4 \otimes \mathbb C^7$ space,  where \(|\varphi_{10}\rangle\), \(|\varphi_{11}\rangle\) and \(|\varphi_{12}\rangle\) cover multiple grids in the figure and thus are not marked separately.  The subsequent figures follow the same convention.}
	\end{widetext}
	
	\vspace*{\fill}
	\clearpage
	\textit{Theorem 1.} \label{Theorem1} In $ \mathbb C^4 \otimes \mathbb C^7$ quantum system, the set $S_1$ consisting of 12 orthogonal quantum states (as shown in Fig. 
\textcolor{blue}{1}) exhibits minimal nonlocality.
	\begin{align}
		|\varphi_1\rangle &= |10\rangle_{AB}-|31\rangle_{AB}, \nonumber \\
		|\varphi_2\rangle &= |20\rangle_{AB}-|32\rangle_{AB}, \nonumber \\
		|\varphi_3\rangle &= |30\rangle_{AB}-|03\rangle_{AB}, \nonumber \\
		|\varphi_4\rangle &= |01\rangle_{AB}-|13\rangle_{AB}, \nonumber \\
		|\varphi_5\rangle &= |02\rangle_{AB}-|23\rangle_{AB}, \nonumber \\
		|\varphi_6\rangle &= |0-1\rangle_{A}|4\rangle_{B}, \nonumber \\
		|\varphi_7\rangle &= |0-1\rangle_{A}|5\rangle_{B}, \nonumber \\
		|\varphi_8\rangle &= |0-1\rangle_{A}|6\rangle_{B}, \nonumber \\
		|\varphi_9\rangle &= |00\rangle_{AB}-|36\rangle_{AB}, \nonumber \\
		|\varphi_{10}\rangle &= |2\rangle_A ( |4\rangle + \omega |5\rangle + \omega^2 |6\rangle)_B, \nonumber \\
		|\varphi_{11}\rangle &= |2\rangle_A ( |4\rangle+ \omega^2 |5\rangle + \omega^4 |6\rangle)_B, \nonumber \\
		|\varphi_{12}\rangle &= \left( \sum_{i=0}^3 |i\rangle \right)_A \left( \sum_{i=0}^6 |i\rangle \right)_B, \nonumber
	\end{align}
	where $\omega = e^{\frac{2\pi \sqrt{-1}}{3}}$.
	
    \begin{proof}
    We first prove that $S_1$ is locally indistinguishable. Suppose Alice performs a measurement first and the measurement preserves orthogonality with a set of POVM elements $\{M_A^\dagger M_A\}$, where each \(M_A^\dagger M_A\) can be written as 
	\[
	M_A^\dagger M_A =
	\begin{bmatrix}
		a_{00} & a_{01} & a_{02} & a_{03} \\
		a_{10} & a_{11} & a_{12} & a_{13} \\
		a_{20} & a_{21} & a_{22} & a_{23} \\
		a_{30} & a_{31} & a_{32} & a_{33}
	\end{bmatrix}_{4\times4}
	\]
	under the basis $\{|0\rangle,\,|1\rangle,\,|2\rangle,\,|3\rangle\}$.
	
	For the measurement to be implementable, the post-measurement states \(\{M_A\otimes I_B|\varphi_i\rangle,\ i = 1, 2, \dots, 12\}\) must be mutually orthogonal. Considering the orthogonal quantum states \(|\varphi_3\rangle\) and \(|\varphi_4\rangle\), we can obtain \(\langle \varphi_3 | M_A^\dagger M_A \otimes I_B^\dagger I_B | \varphi_4 \rangle = 0\) and \(\langle \varphi_4 | M_A^\dagger M_A \otimes I_B^\dagger I_B | \varphi_3 \rangle = 0\), i.e., \((\langle30| - \langle03|\bigr)\bigl(M_A^\dagger M_A \otimes I_B^\dagger I_B\bigr)\bigl(|01\rangle - |13\rangle) = 0\) and \((\langle01| - \langle13|\bigr)\bigl(M_A^\dagger M_A \otimes I_B^\dagger I_B\bigr)\bigl(|30\rangle - |03\rangle) = 0\). Thus, we get \(a_{01} = a_{10} = 0\). Similarly, the remaining nondiagonal entries of the matrix \(M_A^\dagger M_A\) can be calculated to all equal zero, as shown in Table \textcolor{blue}{1}. Considering the orthogonal quantum states \(|\varphi_3\rangle\) and \(|\varphi_{12}\rangle\), we can obtain \(\langle \varphi_3 | M_A^\dagger M_A \otimes I_B^\dagger I_B | \varphi_{12} \rangle = 0\), i.e., $(\langle30| - \langle03|\bigr)\bigl(M_A^\dagger M_A \otimes I_B^\dagger I_B\bigr)\left(\left( \sum_{i=0}^3 |i\rangle \right)_A \left( \sum_{i=0}^6 |i\rangle \right)_B\right) = 0$. Thus, we get \(a_{33}-a_{00} = 0\), i.e., \(a_{33} = a_{00}\). Similarly, we can compute that all remaining diagonal entries of the matrix \(M_A^\dagger M_A\) are identical, as illustrated in Table \textcolor{blue}{2}.
	
	\begin{center}
		\begin{tabular}{c|c} 
			\noalign{\global\arrayrulewidth0.4pt} 
			
			\multicolumn{2}{c}{\textbf{TABLE 1. Nondiagonal elements in $M_A^\dagger M_A$.}} \\
			\hline\hline 
			
			Pair of states & Nondiagonal elements \\
			\hline 
			
			$|\varphi_3\rangle,\ |\varphi_{4}\rangle$ & $a_{10} = a_{01} = 0$ \\
			$|\varphi_3\rangle,\ |\varphi_{5}\rangle$ & $a_{20} = a_{02} = 0$ \\
			$|\varphi_1\rangle,\ |\varphi_{4}\rangle$ & $a_{30} = a_{03} = 0$ \\
			$|\varphi_1\rangle,\ |\varphi_{2}\rangle$ & $a_{12} = a_{21} = 0$ \\
			$|\varphi_1\rangle,\ |\varphi_{3}\rangle$ & $a_{13} = a_{31} = 0$ \\
			$|\varphi_2\rangle,\ |\varphi_{3}\rangle$ & $a_{23} = a_{32} = 0$ \\
			\noalign{\global\arrayrulewidth0.4pt}
			\hline
			\noalign{\global\arrayrulewidth0.4pt}
		\end{tabular}
	\end{center}
	\makeatother
	\label{tab:Nondiagonal elements}
	
	\begin{center}
		\begin{tabular}{c|c} 
			\noalign{\global\arrayrulewidth0.4pt} 
			
			\multicolumn{2}{c}{\textbf{TABLE 2. Diagonal elements in $M_A^\dagger M_A$.}} \\
			\hline\hline 
			
			Pair of states & Diagonal elements \\
			\hline 
			
			$|\varphi_3\rangle,\ |\varphi_{12}\rangle$ & $a_{00} = a_{33} $ \\
			$|\varphi_1\rangle,\ |\varphi_{12}\rangle$ & $a_{11} = a_{33} $ \\
			$|\varphi_2\rangle,\ |\varphi_{12}\rangle$ & $a_{22} = a_{33} $ \\
			\noalign{\global\arrayrulewidth0.4pt}
			\hline
			\noalign{\global\arrayrulewidth0.4pt}
			
		\end{tabular}
	\end{center}
	\makeatother
	\label{tab:diagonal elements}
	
	Tables \textcolor{blue}{1} and \textcolor{blue}{2} show that each $M_A^\dagger M_A$ is proportional to the identity matrix. If Bob performs his measurement first and the measurement preserves orthogonality with a set of POVM elements $\{M_B^\dagger M_B\}$, where each \(M_B^\dagger M_B\) can be written as
	\[
	M_B^\dagger M_B =
	\begin{bmatrix}
		b_{00} & b_{01} & b_{02} & b_{03} & b_{04} & b_{05} & b_{06} \\
		b_{10} & b_{11} & b_{12} & b_{13} & b_{14} & b_{15} & b_{16} \\
		b_{20} & b_{21} & b_{22} & b_{23} & b_{24} & b_{25} & b_{26} \\
		b_{30} & b_{31} & b_{32} & b_{33} & b_{34} & b_{35} & b_{36} \\
		b_{40} & b_{41} & b_{42} & b_{43} & b_{44} & b_{45} & b_{46} \\
		b_{50} & b_{51} & b_{52} & b_{53} & b_{54} & b_{55} & b_{56} \\
		b_{60} & b_{61} & b_{62} & b_{63} & b_{64} & b_{65} & b_{66}
	\end{bmatrix}_{7\times7}
	\]
	under the basis $\{|0\rangle,\,|1\rangle,\,\cdots,\,|6\rangle\}$.
	
	For the measurement to be implementable, the post-measurement states \(\{I_A\otimes M_B \vert\varphi_i\rangle,\ i = 1, 2, \dots, 12\}\) must be mutually orthogonal. Similar to the calculation procedure of Table \textcolor{blue}{1}. Considering the orthogonal quantum states \(|\varphi_4\rangle\) and \(|\varphi_9\rangle\), we can obtain \(\langle \varphi_4 | I_A^\dagger I_A \otimes M_B^\dagger M_B | \varphi_9 \rangle = 0\) and \(\langle \varphi_9 | I_A^\dagger I_A \otimes M_B^\dagger M_B | \varphi_4 \rangle = 0\), i.e., \((\langle01| - \langle13|\bigr)\bigl(I_A^\dagger I_A \otimes M_B^\dagger M_B\bigr)\bigl(|00\rangle - |36\rangle) = 0\) and \((\langle00| - \langle36|\bigr)\bigl(I_A^\dagger I_A \otimes M_B^\dagger M_B\bigr)\bigl(|01\rangle - |13\rangle) = 0\). Thus, we get \(b_{01} = b_{10} = 0\).

		Following this method, we can calculate that all remaining nondiagonal entries equal zero, as shown in Table \textcolor{blue}{3}. It can be seen from Table \textcolor{blue}{3} that all nondiagonal elements of $M_B^\dagger M_B$ are zero. 

    We then prove that the $M_B^\dagger M_B$ are proportional to the identity matrix. Considering the orthogonal states \(|\varphi_3\rangle\) and \(|\varphi_{12}\rangle\), we can obtain 
$\langle \varphi_3 | I_A^\dagger I_A \otimes M_B^\dagger M_B | \varphi_{12} \rangle = 0$, i.e., $(\langle30| - \langle03|\bigr)\bigl(I_A^\dagger I_A \otimes M_B^\dagger M_B)\left(\left( \sum_{i=0}^3 |i\rangle \right)_A \left( \sum_{i=0}^6 |i\rangle \right)_B\right) = 0$. Thus, we get \(b_{33}-b_{00} = 0\), i.e., \(b_{33} = b_{00}\). Similarly, we can obtain \(b_{00}=b_{11}=b_{22}=b_{33}\). Since \(|\varphi_{12}\rangle\) is orthogonal to \(|\varphi_{10}\rangle\) and \(|\varphi_{11}\rangle\) on the second subsystem, we obtain $(\sum_{i=0}^{6}\langle i|) M_B^\dagger M_B \left( |4\rangle + \omega |5\rangle + \omega^2 |6\rangle \right) = 0$ and $(\sum_{i=0}^{6}\langle i|) M_B^\dagger M_B \left( |4\rangle + \omega^2 |5\rangle + \omega^4 |6\rangle \right) = 0$. Thus we have
	
	\begin{align*}
		\begin{cases}
			b_{44} + \omega b_{55} + \omega^2 b_{66} = 0, \\
			b_{44} + \omega^2 b_{55} + \omega^4 b_{66} = 0,
		\end{cases}
		\tag{1}\label{eq:sys1}
	\end{align*}
	where $\omega = e^{\frac{2\pi  \sqrt{-1}}{3}}$.
	
	By Lemma \hyperref[lemma1]{1}, Eq. \eqref{eq:sys1} has a unique solution, i.e., $b_{44}=b_{55}=b_{66}$.
	By $|\varphi_9\rangle$ and $|\varphi_{12}\rangle$, we obtain $b_{00}=b_{66}$.
	Thus, we have $b_{00}=b_{11}=b_{22}=b_{33}=b_{44}=b_{55}=b_{66}$. Thus, any $M_B^\dagger M_B$ is proportional to the identity matrix. Therefore, the set $S_1$ is locally indistinguishable.
\begin{widetext}	
		\begin{center}
			\begin{tabular}{c|c} 
				\noalign{\global\arrayrulewidth0.4pt} 
				
				\multicolumn{2}{c}{\textbf{TABLE 3. Nondiagonal elements in $M_B^\dagger M_B$.}} \\
				\hline\hline 
				
				Pair of states & Nondiagonal elements \\
				\hline 
				
				$|\varphi_4\rangle,\ |\varphi_{9}\rangle$ & $b_{10} = b_{01} = 0$ \\
				$|\varphi_5\rangle,\ |\varphi_{9}\rangle$ & $b_{20} = b_{02} = 0$ \\
				$|\varphi_4\rangle,\ |\varphi_{5}\rangle$ & $b_{12} = b_{21} = 0$ \\
				$|\varphi_3\rangle,\ |\varphi_{4}\rangle$ & $b_{13} = b_{31} = 0$ \\
				$|\varphi_3\rangle,\ |\varphi_{5}\rangle$ & $b_{23} = b_{32} = 0$ \\
				$|\varphi_1\rangle,\ |\varphi_{6}\rangle$ & $b_{40} = b_{04} = 0$ \\
				$|\varphi_1\rangle,\ |\varphi_{7}\rangle$ & $b_{50} = b_{05} = 0$ \\
				$|\varphi_1\rangle,\ |\varphi_{8}\rangle$ & $b_{60} = b_{06} = 0$ \\
				$|\varphi_3\rangle,\ |\varphi_{9}\rangle$ & $b_{30} = b_{06} = 0 ,\ b_{03} = b_{60} = 0$ \\
				$|\varphi_3\rangle,\ |\varphi_{6}\rangle$ & $b_{43} = b_{34} = 0$ \\
				$|\varphi_4\rangle,\ |\varphi_{6}\rangle$ & $b_{14} = b_{34} = 0 ,\ b_{41} = b_{43} = 0 $ \\
				$|\varphi_5\rangle,\ |\varphi_{6}\rangle$ & $b_{42} = b_{43} = 0 ,\ b_{24} = b_{34} = 0$ \\
				$|\varphi_3\rangle,\ |\varphi_{4}\rangle,\ |\varphi_5\rangle,\ |\varphi_{7}\rangle$ & $b_{51} = b_{52} = b_{53} = b_{35} = b_{25} = b_{15} =  0$ \\
				$|\varphi_3\rangle,\ |\varphi_{4}\rangle,\ |\varphi_5\rangle,\ |\varphi_{8}\rangle$ & $b_{61} = b_{62} = b_{63} = b_{36} = b_{26} = b_{16} =  0$ \\
				$|\varphi_6\rangle,\ |\varphi_{7}\rangle$ & $b_{45} = b_{54} = 0$ \\
				$|\varphi_6\rangle,\ |\varphi_{8}\rangle$ & $b_{46} = b_{64} = 0$ \\
				$|\varphi_7\rangle,\ |\varphi_{8}\rangle$ & $b_{56} = b_{65} = 0$\\
				\noalign{\global\arrayrulewidth0.4pt}
				\hline
				\noalign{\global\arrayrulewidth0.4pt}
			\end{tabular}
		\end{center}
		\makeatother
		\label{tab:diagonal_elements}
	\end{widetext}
	
	Next, we prove that $S_1 - \{|\varphi_{12}\rangle\}$ is distinguishable under LOCC.
	As the first step, Alice performs a measurement with the operators: $\{A_i = |i\rangle_A\langle i|\}$, where $0\le i\le 3$.
	
	(1) If Alice’s measurement outcome corresponds to $A_0 = |0\rangle_A \langle 0|$, the measured state must be one of
	$\{|\varphi_{i}\rangle \mid i=3,\ 4,\ \dots,\ 9\}$, and it will collapse to one of the following forms:
	
	\begin{align}
		|\varphi_3\rangle &= |30\rangle_{AB}-|03\rangle_{AB} \to |03\rangle_{AB}, \nonumber \\
		|\varphi_4\rangle &= |01\rangle_{AB}-|13\rangle_{AB} \to |01\rangle_{AB}, \nonumber \\
		|\varphi_5\rangle &= |02\rangle_{AB}-|23\rangle_{AB} \to |02\rangle_{AB}, \nonumber \\
		|\varphi_6\rangle &= |0-1\rangle_{A}|4\rangle_{B} \to |04\rangle_{AB}, \nonumber \\
		|\varphi_7\rangle &= |0-1\rangle_{A}|5\rangle_{B} \to |05\rangle_{AB}, \nonumber \\
		|\varphi_8\rangle &= |0-1\rangle_{A}|6\rangle_{B} \to |06\rangle_{AB}, \nonumber \\
		|\varphi_9\rangle &= |00\rangle_{AB}-|36\rangle_{AB} \to |00\rangle_{AB}. \nonumber 
	\end{align}
	
	Since states $\{|03\rangle,\,|01\rangle,\,|02\rangle,\,|04\rangle,\,|05\rangle,\,|06\rangle\}$ are mutually orthogonal on Bob’s subsystem, they can be perfectly distinguished by Bob.
	
	(2) If Alice’s measurement outcome corresponds to $A_1 = |1\rangle_A \langle 1|$, the measured state must be one of $\{|\varphi_1\rangle,\,|\varphi_4\rangle,\,|\varphi_6\rangle,\,|\varphi_7\rangle,\,|\varphi_8\rangle\}$, and it will collapse to one of the following forms:
	
	\begin{align}
		|\varphi_1\rangle &= |10\rangle_{AB}-|31\rangle_{AB} \to |10\rangle_{AB}, \nonumber \\
		|\varphi_4\rangle &= |01\rangle_{AB}-|13\rangle_{AB} \to |13\rangle_{AB}, \nonumber \\
		|\varphi_6\rangle &= |0-1\rangle_{A}|4\rangle_{B} \to |14\rangle_{AB}, \nonumber \\
		|\varphi_7\rangle &= |0-1\rangle_{A}|5\rangle_{B} \to |15\rangle_{AB}, \nonumber \\
		|\varphi_8\rangle &= |0-1\rangle_{A}|6\rangle_{B} \to |16\rangle_{AB}. \nonumber 
	\end{align}
	
	Since states \(\{|10\rangle,\,|13\rangle,\,|14\rangle,\,|15\rangle,\,|16\rangle\}\) are mutually orthogonal on Bob’s subsystem, they can be perfectly distinguished by Bob.
	
	(3) If Alice’s measurement outcome corresponds to $A_2 = |2\rangle_A \langle 2|$, the measured state must be one of $\{|\varphi_2\rangle,\,|\varphi_5\rangle,\,|\varphi_{10}\rangle,\,|\varphi_{11}\rangle\}$, and it will collapse to one of the following forms:
	\begin{align}
		|\varphi_2\rangle &= |20\rangle_{AB}-|32\rangle_{AB} \to |20\rangle_{AB}, \nonumber \\
		|\varphi_5\rangle &= |02\rangle_{AB}-|23\rangle_{AB} \to |23\rangle_{AB}, \nonumber \\
		|\varphi_{10}\rangle &= |2\rangle_A ( |4\rangle + \omega |5\rangle + \omega^2 |6\rangle)_B, \nonumber \\ 
		|\varphi_{11}\rangle &= |2\rangle_A (|4\rangle + \omega^2 |5\rangle+ \omega^4 |6\rangle)_B. \nonumber 
	\end{align}
	
	Similarly, the collapsed states are mutually orthogonal on Bob’s subsystem, so they can be perfectly distinguished by Bob.
	
	(4) If Alice’s measurement outcome corresponds to $A_3 = |3\rangle_A \langle 3|$, the measured state must be one of
	$\{|\varphi_{1}\rangle,\allowbreak|\varphi_{2}\rangle,\allowbreak|\varphi_{3}\rangle,\allowbreak|\varphi_{9}\rangle\}$, and it will collapse to one of the following forms:
	
	\begin{align}
		|\varphi_1\rangle &= |10\rangle_{AB}-|31\rangle_{AB} \to |31\rangle_{AB}, \nonumber \\
		|\varphi_2\rangle &= |20\rangle_{AB}-|32\rangle_{AB} \to |32\rangle_{AB}, \nonumber \\
		|\varphi_3\rangle &= |30\rangle_{AB}-|03\rangle_{AB} \to |30\rangle_{AB}, \nonumber \\
		|\varphi_9\rangle &= |00\rangle_{AB}-|36\rangle_{AB} \to |36\rangle_{AB}. \nonumber 
	\end{align}
	
	Since states \(\{|31\rangle,\,|32\rangle,\,|30\rangle,\,|36\rangle\}\) are mutually orthogonal on Bob’s subsystem, they can be perfectly distinguished by Bob.
	
		In summary, the set $S_1 - \{|\varphi_{12}\rangle\}$ can be locally distinguished.  This completes the proof.\end{proof}

        Now, we give a method to construct a set of orthogonal quantum states with minimal nonlocality in bipartite quantum system of of unequal local dimensions. 
		
		\textit{Theorem 2.} \label{Theorem2} In $ \mathbb C^{d_1} \otimes \mathbb C^{d_2}$ quantum system, the following set $S_2$ consisting of $2d_2-2$ orthogonal quantum states (as shown in Fig. \textcolor{blue}{2}) exhibits minimal nonlocality, where $4\le d_1 < d_2$.    
\begin{widetext}	
		\begin{align*}
			|\varphi_i\rangle &= |i0\rangle_{AB} - |(d_1-1)i\rangle_{AB}, & 1 &\le i \le d_1-2,\nonumber  \\
			|\varphi_{d_1-1}\rangle &= |(d_1-1)0\rangle_{AB} - |0(d_1-1)\rangle_{AB}, \nonumber \\
			|\varphi_{i+d_1-1}\rangle &= |0i\rangle_{AB} - |i(d_1-1)\rangle_{AB}, & 1 &\le i \le d_1-2,\nonumber  \\
			|\varphi_{2d_1-3+j}\rangle &= |0-1\rangle_A |(d_1-1+j)\rangle_B, & 1 &\le j \le d_2-d_1, \nonumber \\
			|\varphi_{d_1+d_2-2}\rangle &= |00\rangle_{AB} - |(d_1-1)(d_2-1)\rangle_{AB} ,\nonumber \\
			|\varphi_{d_1+d_2-2+i}\rangle &= |2\rangle_A \left(\sum_{j=0}^{d_{2}-d_{1}-1}\omega^{ij}|(d_{1}+j)\rangle\right)_B, & 1 &\le i \le d_2-d_1-1, \nonumber \\
			|\varphi_{2d_2-2}\rangle &= \left(\sum_{k=0}^{d_1-1} |k\rangle\right)_A \left(\sum_{k=0}^{d_2-1} |k\rangle\right)_B, \nonumber 
		\end{align*}
\begin{center}
			\begin{tikzpicture}[
				box/.style = {
					draw,
					line width=0.7mm,
					minimum size=1.6cm,
					font=\scriptsize\bfseries
				},
				c1/.style = {fill=red!25},
				c2/.style = {fill=blue!15},
				c3/.style = {fill=green!15},
				c4/.style = {fill=orange!15},
				c5/.style = {fill=purple!25},
				c6/.style = {fill=cyan!15},
				c7/.style = {fill=pink!15},
				c8/.style = {fill=yellow!25},
				c9/.style = {fill=gray!25}
				]
				
				\node at (0, 1.2) {$|0\rangle$};
				\node at (1.6, 1.2) {$|1\rangle$};
				\node at (3.2, 1.2) {$\cdots$};
				\node at (4.8, 1.2) {$|{(d_1-2)}\rangle$};
				\node at (6.4, 1.2) {$|{(d_1-1)}\rangle$};
				\node at (8.0, 1.2) {$|{d_1}\rangle$};
				\node at (9.6, 1.2) {$\cdots$};
				\node at (11.2, 1.2) {$|{(d_2-1)}\rangle$};
				
				\node at (-1.6, 0) {$|0\rangle$};
				\node at (-1.6, -1.6) {$|1\rangle$};
				\node at (-1.6, -3.2) {$\vdots$};
				\node at (-1.6, -4.8) {$|{(d_1-2)}\rangle$};
				\node at (-1.6, -6.4) {$|{(d_1-1)}\rangle$};
				
				\node[box, c9] at (0, 0)    {$d_1+d_2-2$};
				\node[box, c4] at (1.6, 0)  {$d_1$};
				\node[box] at (3.2, 0)      {$\cdots$};
				\node[box, c3] at (4.8, 0)  {$2d_1-3$};
				\node[box, c6] at (6.4, 0)  {$d_1-1$};
				\node[box, c7] at (8.0, 0)  {$2d_1-2$};
				\node[box] at (9.6, 0)      {$\cdots$};
				\node[box, c8] at (11.2, 0) {$d_1+d_2-3$};
				
				\node[box, c1] at (0, -1.6)    {$1$};
				\node[box] at (1.6, -1.6)      {};
				\node[box] at (3.2, -1.6)      {};
				\node[box] at (4.8, -1.6)      {};
				\node[box, c4] at (6.4, -1.6)  {$d_1$};
				\node[box, c7] at (8.0, -1.6)  {$2d_1-2$};
				\node[box] at (9.6, -1.6)      {$\cdots$};
				\node[box, c8] at (11.2, -1.6) {$d_1+d_2-3$};
				
				\node[box] at (0, -3.2)    {$\vdots$};
				\node[box] at (1.6, -3.2)  {};
				\node[box] at (3.2, -3.2)  {};
				\node[box] at (4.8, -3.2)  {};
				\node[box] at (6.4, -3.2)  {$\vdots$};
				\node[box] at (8.0, -3.2)  {};
				\node[box] at (9.6, -3.2)  {};
				\node[box] at (11.2, -3.2) {};
				
				\node[box, c2] at (0, -4.8)    {$d_1-2$};
				\node[box] at (1.6, -4.8)      {};
				\node[box] at (3.2, -4.8)      {};
				\node[box] at (4.8, -4.8)      {};
				\node[box, c3] at (6.4, -4.8)  {$2d_1-3$};
				\node[box] at (8.0, -4.8)      {};
				\node[box] at (9.6, -4.8)      {};
				\node[box] at (11.2, -4.8)     {};
				
				\node[box, c6] at (0, -6.4)    {$d_1-1$};
				\node[box, c1] at (1.6, -6.4)  {$1$};
				\node[box] at (3.2, -6.4)      {$\cdots$};
				\node[box, c2] at (4.8, -6.4)  {$d_1-2$};
				\node[box] at (6.4, -6.4)      {};
				\node[box] at (8.0, -6.4)      {};
				\node[box] at (9.6, -6.4)      {};
				\node[box, c9] at (11.2, -6.4) {$d_1+d_2-2$};
				
				\node[font=\bfseries] at (5.6, 1.8) {Bob};
				\node[font=\bfseries, rotate=90] at (-2, -3.2) {Alice};
				
			\end{tikzpicture}
			
			\mycaption{The structure of the set $S_2$ in $ \mathbb C^{d_1} \otimes \mathbb C^{d_2}$ space,  where $4\le d_1< d_2$.}
		\end{center}
\end{widetext}	
where $\omega = e^{\frac{2\pi  \sqrt{-1}}{d_2-d_1}}$.

		 \begin{proof} Suppose Alice first performs an orthogonality-preserving measurement with a set of POVM elements $\{M_A^\dagger M_A\}$, where each \(M_A^\dagger M_A\) can be written as  
		\[
		M_A^\dagger M_A =
		\begin{bmatrix}
			a_{00} & a_{01} & \cdots & a_{0(d_1-1)} \\
			a_{10} & a_{11} & \cdots & a_{1(d_1-1)} \\
			\vdots & \vdots & \ddots & \vdots \\
			a_{(d_1-1)0} & a_{(d_1-1)1} & \cdots & a_{(d_1-1)(d_1-1)}
		\end{bmatrix}_{d_1\times d_1}
		\]
		under the basis $\{|0\rangle,\,|1\rangle,\,\cdots,\,|(d_1-1)\rangle\}$.
		
		For the measurement to be implementable, the post-measurement states $\{ M_{A} \otimes I_{B}|\varphi_{i}\rangle \mid i=1, $ $2,$ $\dots,$ $2d_{2}-2\}$ must be mutually orthogonal. Considering the states \(|\varphi_i\rangle\) and \(|\varphi_k\rangle\), \(1 \le i \neq k \le d_1-2\), we can obtain \(\langle \varphi_i| M_A^\dagger M_A \otimes I_B^\dagger I_B| \varphi_k \rangle = 0\) i.e., \([\langle i | M_A^\dagger M_A | k \rangle][\langle 0 | I_B^\dagger I_B | 0 \rangle] - [\langle i | M_A^\dagger M_A | (d_1-1) \rangle][\langle 0 | I_B^\dagger I_B | k \rangle] - [\langle (d_1-1) | M_A^\dagger M_A | k \rangle][\langle i | I_B^\dagger I_B | 0 \rangle] + [\langle (d_1-1) | M_A^\dagger M_A |(d_1-1) \rangle][\langle i | I_B^\dagger I_B | k \rangle]  = 0\). Thus, we get \(a_{ik} =0\). Similarly, we can also obtain \(a_{ki} =0\) by \(\langle \varphi_k| M_A^\dagger M_A \otimes I_B^\dagger I_B| \varphi_i \rangle = 0\). Following this method, we can calculate that all remaining nondiagonal entries equal zero, as shown in Table \textcolor{blue}{4}. For diagonal elements, we can calculate and conclude that all diagonal elements are equal, as shown in Table \textcolor{blue}{5}.
\begin{widetext}	
		
		\begin{center}
			\begin{tabular}{c|c|c} 
				
				\multicolumn{3}{c}{\textbf{TABLE 4. Nondiagonal elements in $M_A^\dagger M_A$.}} \\
				\hline\hline
				
				Pair of states & Noniagonal elements & Ranges \\
				\hline
				
				$|\varphi_i\rangle,\ |\varphi_{k}\rangle$ & $a_{ik} = a_{ki} =0$ & $1 \le i \neq k \le d_1-2$ \\
				$|\varphi_i\rangle,\ |\varphi_{d_1-1}\rangle$ & $a_{i(d_1-1)} = a_{(d_1-1)i} =0$ & $1 \le i \le d_1-2$ \\
				$|\varphi_{d_1-1}\rangle,\ |\varphi_{i+d_1-1}\rangle$ & $a_{i0} = a_{0i} = 0$ & $1 \le i \le d_1-2$ \\
				$|\varphi_{i}\rangle,\ |\varphi_{i+d_1-1}\rangle$ & $a_{(d_1-1)0} = a_{0(d_1-1)} = 0$ & $-$ \\
				\hline
			\end{tabular}
		\end{center}
		\label{tab:Nondiagonal_elements}
		
		\begin{center}
			\begin{tabular}{c|c|c} 
				
				\multicolumn{3}{c}{\textbf{TABLE 5. Diagonal elements in $M_A^\dagger M_A$.}} \\
				\hline\hline
				
				Pair of states & Diagonal elements & Ranges \\
				\hline
				
				$|\varphi_i\rangle,\ |\varphi_{2d_2-2}\rangle$ & $a_{ii} = a_{(d_1-1)(d_1-1)}$ & $1 \le i \le d_1-2$ \\
				$|\varphi_{i+d_1-1}\rangle,\ |\varphi_{2d_2-2}\rangle$ & $a_{00} = a_{ii} $ & $1 \le i \le d_1-2$ \\
				\hline
			\end{tabular}
		\end{center}
		\label{tab:Diagonal_elements}
\begin{center}
			\begin{tabular}{c|c|c} 
				
				\multicolumn{3}{c}{\textbf{TABLE 6. Nondiagonal elements in $M_B^\dagger M_B$.}} \\
				\hline\hline
				
				Pair of states & Nondiagonal elements & Ranges \\
				\hline
				
				$|\varphi_{i+d_1-1}\rangle,\ |\varphi_{k+d_1-1}\rangle$ & $b_{ik} = b_{ki} =0$ & $1\le i \neq k \le d_1-2$ \\
				$|\varphi_{d_1-1}\rangle,\ |\varphi_{i+d_1-1}\rangle$ & $b_{(d_1-1)i} = b_{i(d_1-1)} =0$ & $1\le i\le d_1-2$ \\
				$|\varphi_i\rangle,\ |\varphi_{d_1-1}\rangle$ & $b_{i0} = b_{0i} =0$ & $1 \le i \le d_1-2$ \\
				$|\varphi_{d_1-1}\rangle,\ |\varphi_{2d_1-3+j}\rangle$ & $b_{(d_1-1)(d_1-1+j)} = b_{(d_1-1+j)(d_1-1)} =0$ & $1 \le j \le d_2-d_1$ \\
				$|\varphi_{i+d_1-1}\rangle,\ |\varphi_{2d_1-3+j}\rangle$ & $b_{i(d_1-1+j)} = b_{(d_1-1+j)i} =0$ & $1 \le j \le d_2-d_1 ,1 \le i \le d_1-2$ \\
				$|\varphi_{2d_1-3+j}\rangle,\ |\varphi_{2d_1-3+k}\rangle$ & $b_{(d_1-1+k)(d_1-1+j)} = b_{(d_1-1+j)(d_1-1+k)} = 0$ & $1 \le j \neq k \le d_2-d_1$ \\
				$|\varphi_{1}\rangle,\ |\varphi_{2d_1-3+j}\rangle$ & $b_{(d_1-1+j)0} = b_{0(d_1-1+j)} = 0$ & $1 \le j \le d_2-d_1$ \\
				$|\varphi_{d_1-1}\rangle,\ |\varphi_{d_1+d_2-2}\rangle$ & $b_{0(d_2-1)} = b_{(d_1-1)0} = 0,\ b_{(d_2-1)0} = b_{0(d_1-1)} = 0$ & $-$ \\
				\hline
			\end{tabular}
		\end{center}
		\label{tab:Nondiagonal_elements}
	\end{widetext}
		
		From Tables \textcolor{blue}{4} and \textcolor{blue}{5}, we see that any POVM element $M_A^\dagger M_A$ is proportional to the identity matrix. Therefore, Alice can only perform a trivial measurement. 

           On the other hand, suppose Bob first performs an orthogonality-preserving measurement with a set of POVM elements $\{M_B^\dagger M_B\}$, where each \(M_B^\dagger M_B\) can be written as
		\[
		M_B^\dagger M_B =
		\begin{bmatrix}
			b_{00} & b_{01} & \cdots & b_{0(d_2-1)} \\
			b_{10} & b_{11} & \cdots & b_{1(d_2-1)} \\
			\vdots & \vdots & \ddots & \vdots \\
			b_{(d_2-1)0} & b_{(d_2-1)1} & \cdots & b_{(d_2-1)(d_2-1)}
		\end{bmatrix}_{d_2\times d_2}
		\]
		under the basis $\{|0\rangle,\,|1\rangle,\,\cdots,\,|(d_2-1)\rangle\}$.
		
		Similar to the previous proof, Bob performs a measurement using the orthogonal-preserving POVM elements $M_B^\dagger M_B$, and the results are listed in Table \textcolor{blue}{6}. It follows from Table \textcolor{blue}{6} that all nondiagonal elements of $M_B^\dagger M_B$ are zero. 

        We then prove that the diagonal elements of $M_B^\dagger M_B$ are equal. Since $|\varphi_{2d_2-2}\rangle$ is orthogonal to $|\varphi_{i}\rangle$ and $|\varphi_{i+d_1-1}\rangle$ on Bob's subsystem, we obtain $b_{00}=b_{11}=\dots=b_{(d_1-1)(d_1-1)}$. Furthermore, \(|\varphi_{d_1+d_2-2+i}\rangle\) is also orthogonal to \(|\varphi_{2d_2-2}\rangle\) on Bob's subsystem, we derive the following system of equations:
		
		\begin{align*}
			\begin{cases}
			  \sum_{j=0}^{d_{2}-d_{1}-1}\omega^{j}b_{(d_{1}+j)(d_{1}+j)}=0,\\
               \sum_{j=0}^{d_{2}-d_{1}-1}\omega^{2j}b_{(d_{1}+j)(d_{1}+j)}=0,\\
				\quad\quad\quad\quad\quad\quad\vdots\\[4pt]
             \sum_{j=0}^{d_{2}-d_{1}-1}\omega^{(d_2-d_1-1)j}b_{(d_{1}+j)(d_{1}+j)}=0,\\ 
				\end{cases}
			\tag{2}\label{eq:sys2}
		\end{align*}
		where $\omega = e^{\frac{2\pi \sqrt{-1}}{d_2-d_1}}$.
		
		By Lemma \hyperref[lemma1]{1}, this system of Eq. \eqref{eq:sys2} has a unique solution, namely
		$b_{d_1d_1} = b_{(d_1+1)(d_1+1)} = \cdots = b_{(d_2-1)(d_2-1)}$. From \(|\varphi_{d_1+d_2-2}\rangle\) and \(|\varphi_{2d_2-2}\rangle\), we obtain \(b_{00}=b_{(d_2-1)(d_2-1)}\). Thus, we have proven that all diagonal elements are equal. Therefore, $S_2$ cannot be distinguished by LOCC.
		
		Next, we prove that $S_2 - \{|\varphi_{2d_2-2}\rangle\}$ is distinguishable under LOCC. As the first step, Alice performs measurement with operators: $\{A_0 = |0\rangle_A\langle 0|\ ,\, A_1 = |1\rangle_A\langle 1|\ , \,A_2 = |2\rangle_A\langle 2|\ , \,A_3 = |(d_1-1)\rangle_A\langle(d_1-1)|\ , \,A_4 = I_A-A_0-A_1-A_2-A_3\}$ .
		
		(1) If Alice’s measurement outcome corresponds to $A_0 = |0\rangle_A \langle 0|$, the state measured must be one of
		$\{|\varphi_{d_1-1}\rangle,\allowbreak|\varphi_{i+d_1-1}\rangle,\allowbreak|\varphi_{2d_1-3+j}\rangle,\allowbreak|\varphi_{d_1+d_2-2}\rangle\}$, and it will collapse to one of the following forms:

		\begin{align}
			|\varphi_{d_1-1}\rangle & \to |0(d_1-1)\rangle_{AB}, \nonumber \\
			|\varphi_{i+d_1-1}\rangle &\to |0i\rangle_{AB}, \quad 1 \le i \le d_1-2, \nonumber\\
			|\varphi_{2d_1-3+j}\rangle &\to |0(d_1-1+j)\rangle_{AB}, \quad 1\le j\le d_2-d_1, \nonumber\\
			|\varphi_{d_1+d_2-2}\rangle &\to |00\rangle_{AB}.\nonumber 
		\end{align}

		Since $d_1\le j\le d_2-1$, we have $d_1\le d_1-1+j\le d_2-1$. Thus the collapsed states are mutually orthogonal on Bob's subsystem, which implies that $|\varphi_{d_1-1}\rangle$, $|\varphi_{i+d_1-1}\rangle$, $|\varphi_{2d_1-3+j}\rangle$ and $|\varphi_{d_1+d_2-2}\rangle$ can be distinguished via LOCC.
		
		(2) If Alice’s measurement outcome corresponds to $A_1 = |1\rangle_A \langle 1|$, the measured state must be one of
		$\{|\varphi_{1}\rangle,\,\allowbreak|\varphi_{d_1}\rangle,\,\allowbreak|\varphi_{2d_1-3+j}\rangle\}$, and it will collapse to one of the following forms:
	
		\begin{align}
			|\varphi_{1}\rangle & \to |10\rangle_{AB}, \nonumber \\
			|\varphi_{d_1}\rangle &\to |1(d_1-1)\rangle_{AB}, \nonumber\\
			|\varphi_{2d_1-3+j}\rangle & \to |1(d_1-1+j)\rangle_{AB}, \quad 1\le j\le d_2-d_1. \nonumber 
		\end{align}
		
		Obviously, the collapsed states are mutually orthogonal on Bob's subsystem. Hence, $|\varphi_{1}\rangle$, $|\varphi_{d_1}\rangle$ and $|\varphi_{2d_1-3+j}\rangle$ can also be distinguished via LOCC.
		
		(3) If Alice’s measurement outcome corresponds to $A_2 = |2\rangle_A \langle 2|$, the measured state must be one of
		$\{|\varphi_{2}\rangle,\,\allowbreak|\varphi_{d_1+1}\rangle,\,\allowbreak|\varphi_{d_1+d_2-2+i}\rangle\}$, and it will collapse to one of the following forms:
	
		\begin{align}
			|\varphi_{2}\rangle & \to |20\rangle_{AB}, \nonumber \\
			|\varphi_{d_1+1}\rangle &\to |2(d_1-1)\rangle_{AB}, \nonumber\\
            |\varphi_{d_1+d_2-2+i}\rangle & \to |\varphi_{d_1+d_2-2+i}\rangle, & 1 &\le i \le d_2-d_1-1, \nonumber 
			\end{align}
		where $\omega = e^{\frac{2\pi  \sqrt{-1}}{d_2-d_1}}$
		
		By the properties of roots of unity, $|\varphi_{d_1+d_2-2+i}\rangle$ for distinct $i$ are mutually orthogonal on Bob's subsystem. The collapsed components of the other two states on Bob's subsystem correspond to $0$ and $d_1-1$, respectively. Consequently, all remaining states are orthogonal on Bob's subsystem, so $|\varphi_{2}\rangle$, $|\varphi_{d_1+1}\rangle$ and $|\varphi_{d_1+d_2-2+i}\rangle$ can be distinguished via LOCC.
		
		(4) If Alice’s measurement outcome corresponds to $A_3 = |(d_1-1)\rangle_A\langle(d_1-1)|$, the measured state must be one of $\{|\varphi_{i}\rangle,\,\allowbreak|\varphi_{d_1-1}\rangle,\,\allowbreak|\varphi_{d_1+d_2-2}\rangle\}$, it will collapse to one of the following forms:

		\begin{align}
			|\varphi_{i}\rangle &\to |(d_1-1)i\rangle_{AB},&& 1\le i\le d_1-2, \nonumber \\
			|\varphi_{d_1-1}\rangle &\to |(d_1-1)0\rangle_{AB}, \nonumber\\
			|\varphi_{d_1+d_2-2}\rangle & \to |(d_1-1)(d_2-1)\rangle_{AB}. \nonumber 
		\end{align}
		
		Obviously, the collapsed states are mutually orthogonal on Bob's subsystem. Hence, $|\varphi_{i}\rangle,\,\allowbreak|\varphi_{d_1-1}\rangle,\,\allowbreak|\varphi_{d_1+d_2-2}\rangle$ can also be distinguished via LOCC.
		
		(5) If Alice’s measurement outcome corresponds to $A_4 = I_A-A_0-A_1-A_2-A_3$, the measured state must be one of
		$\{|\varphi_{i}\rangle,\,\allowbreak|\varphi_{i+d_1-1}\rangle \mid 3 \le i \le d_1-2\}$, and it will collapse to one of the following forms:

		\begin{align}
			|\varphi_{i}\rangle & \to |i0\rangle_{AB},&& 3\le i\le d_1-2 \nonumber \\
			|\varphi_{i+d_1-1}\rangle & \to |i(d_1-1)\rangle_{AB}, && 3\le i\le d_1-2. \nonumber 
		\end{align}
	
	Similarly, the remaining states are still orthogonal on Bob's subsystem. Thus $|\varphi_{i}\rangle$ and $|\varphi_{i+d_1-1}\rangle$ with $3 \le i \le d_1-2$ can be distinguished via LOCC.
	
	In summary, $S_2 - \{|\varphi_{2d_2-2}\rangle\}$ is distinguishable via LOCC. Therefore, $S_2$ possesses minimal nonlocality. 
 \end{proof}
	
	\section{Minimal Nonlocality of Orthogonal Quantum State Set in a tripartite unequal-dimensional quantum system}
	\label{sec:4}
	In this section, we give a method to construct a set of ststes with minimal nonlocality in a tripartite unequal-dimensional quantum system. And then we give a proof for minimal nonlocality of the set.
	
		\textit{Theorem 3.} \label{Theorem3} In $ \mathbb C^{d_1} \otimes \mathbb C^{d_2}\otimes \mathbb C^{d_3}$ quantum system, the following set $S_3$ consisting of $4d_3+2d_1-11$ orthogonal quantum states exhibits minimal nonlocality, where $5\le d_1 < d_2 < d_3$.  
\begin{widetext}
		\begin{align*}
			|\varphi_i\rangle &= |(i+1)0(d_3-2)\rangle_{ABC} - |(d_1-1)0(d_3-2-i)\rangle_{ABC}, & 1 &\le i \le d_1-3,\nonumber  \\
			|\varphi_{d_1-2}\rangle &= |(d_1-1)0(d_3-2)\rangle_{ABC} - |10(d_3-d_1)\rangle_{ABC}, \nonumber \\
			|\varphi_{i+d_1-2}\rangle &= |10(d_3-2-i)\rangle_{ABC} - |(1+i)0(d_3-d_1)\rangle_{ABC}, & 1 &\le i \le d_1-3,\nonumber  \\
			|\varphi_{2d_1-5+i}\rangle &= |1-2\rangle_A |0\rangle_B|(d_3-d_1-i)\rangle_C, & 1 &\le i \le d_3-d_1, \nonumber \\
			|\varphi_{d_3+d_1-4}\rangle &= |(d_1-1)00\rangle_{ABC} - |10(d_3-2)\rangle_{ABC} ,\nonumber \\
			|\varphi_{d_3+d_1-4+i}\rangle &= |0(d_2-2-i)1\rangle_{ABC} - |00(1+i)\rangle_{ABC}, & 1 &\le i \le d_2-3,\nonumber  \\
			|\varphi_{d_3+d_2+d_1-6}\rangle &= |001\rangle_{ABC} - |0(d_2-2)(d_2-1)\rangle_{ABC}, \nonumber \\
			|\varphi_{i+d_3+d_2+d_1-6}\rangle &= |0(d_2-2)(1+i)\rangle_{ABC} - |0(d_2-2-i)(d_2-1)\rangle_{ABC}, & 1 &\le i \le d_2-3,\nonumber  \\
			|\varphi_{d_3+2d_2+d_1-9+i}\rangle &= |0\rangle_A |(d_2-2)-(d_2-3)\rangle_B|(d_2-1+i)\rangle_C, & 1 &\le i \le d_3-d_2, \nonumber \\
			|\varphi_{2d_3+d_2+d_1-8}\rangle &= |0(d_2-2)1\rangle_{ABC} - |00(d_3-1)\rangle_{ABC} ,\nonumber \\
			|\varphi_{2d_3+d_2+d_1-8+i}\rangle &= |0\rangle_A|(d_2-4)\rangle_B \left(\sum_{j=0}^{d_3-d_2-1}\omega^{ji}|(d_2+j)\rangle\right)_C, & 1 &\le i \le d_3-d_2-1, \nonumber \\
			|\varphi_{3d_3+d_1-9+i}\rangle &= |(d_1-2-i)10\rangle_{ABC} - |0(1+i)0\rangle_{ABC}, & 1 &\le i \le d_1-3,\nonumber  \\
			|\varphi_{3d_3+2d_1-11}\rangle &= |010\rangle_{ABC} - |(d_1-2)(d_1-1)0\rangle_{ABC}, \nonumber \\
			|\varphi_{i+3d_3+2d_1-11}\rangle &= |(d_1-2)(1+i)0\rangle_{ABC} - |(d_1-2-i)(d_1-1)0\rangle_{ABC}, & 1 &\le i \le d_1-3,\nonumber  \\
			|\varphi_{3d_3+2d_2+d_1-14+i}\rangle &= |(d_1-2)-(d_1-3)\rangle_A |(d_1-1+i)\rangle_B|0\rangle_C, & 1 &\le i \le d_2-d_1, \nonumber \\
			|\varphi_{3d_3+d_2+2d_1-13}\rangle &= |(d_1-2)10\rangle_{ABC} - |0(d_2-1)0\rangle_{ABC} ,\nonumber \\
			|\varphi_{3d_3+d_2+2d_1-12}\rangle &= |000\rangle_{ABC} - |111\rangle_{ABC}, \nonumber\\
			|\varphi_{3d_3+d_2+2d_1-12+i}\rangle &= |0\rangle_A |0-(d_2-1)\rangle_B|(d_2-2+i)\rangle_C,& 1 &\le i \le d_3-d_2, \nonumber\\
			|\varphi_{4d_3+2d_1-11}\rangle &= \left(\sum_{k=0}^{d_1-1} |k\rangle\right)_A \left(\sum_{k=0}^{d_2-1} |k\rangle\right)_B \left(\sum_{k=0}^{d_3-1} |k\rangle\right)_C, \nonumber
		\end{align*}
       where  $\omega = e^{\frac{2\pi \sqrt{-1}}{d_3-d_2}}$.    
    \end{widetext}
	   \begin{proof}
	   	\begin{widetext}
       We first prove that $S_3$ is locally indistinguishable. Suppose Alice first performs an orthogonality-preserving measurement with a set of POVM elements $\{M_A^\dagger M_A\}$, 
 where each \(M_A^\dagger M_A\) can be written as  

		\[
		M_A^\dagger M_A =
		\begin{bmatrix}
			a_{00} & a_{01} & \cdots & a_{0(d_1-1)} \\
			a_{10} & a_{11} & \cdots & a_{1(d_1-1)} \\
			\vdots & \vdots & \ddots & \vdots \\
			a_{(d_1-1)0} & a_{(d_1-1)1} & \cdots & a_{(d_1-1)(d_1-1)}
		\end{bmatrix}_{d_1\times d_1}
		\]
		under the basis $\{|0\rangle,\,|1\rangle,\,\cdots,\,|(d_1-1)\rangle\}$.
		
		Similar to the proof of Theorem \hyperref[Theorem2]{2}, Alice carries out an initial measurement via orthogonality-preserving POVM elements \(\{M_A^\dagger M_A\}\), with the results summarized in Tables \textcolor{blue}{7} and \textcolor{blue}{8}.
		
		\begin{center}
			\begin{tabular}{c|c|c} 
				
				\multicolumn{3}{c}{\textbf{TABLE 7. Nondiagonal elements in $M_A^\dagger M_A$.}} \\
				\hline\hline
				
				Pair of states & Nondiagonal elements & Ranges \\
				\hline
				
				$|\varphi_i\rangle,\ |\varphi_{d_1-2}\rangle$ & $a_{(d_1-1)(i+1)} = a_{(i+1)(d_1-1)} =0$ & $1 \le i \le d_1-3$ \\
				$|\varphi_i\rangle,\ |\varphi_{j}\rangle$ & $a_{(i+1)(j+1)} = a_{(j+1)(i+1)} =0$ & $1 \le i \neq j \le d_1-3$ \\ 
				$|\varphi_{i}\rangle,\ |\varphi_{i+d_1-2}\rangle$ & $a_{(d_1-1)1} = a_{1(d_1-1)} = 0$ & $-$ \\
				$|\varphi_{d_1-2}\rangle,\ |\varphi_{i+d_1-2}\rangle$ & $a_{(i+1)1} = a_{1(i+1)} = 0$ & $1 \le i \le d_1-3$ \\
				$|\varphi_{3d_3+d_1-9+i}\rangle,\ |\varphi_{3d_3+2d_1-11}\rangle$ & $a_{0(d_1-2-i)} = a_{(d_1-2-i)0} = 0$ & $1 \le i \le d_1-3$ \\
				$|\varphi_{3d_3+d_1-9+i}\rangle,\ |\varphi_{i+3d_3+2d_1-11}\rangle$ & $a_{0(d_1-2)} = a_{(d_1-2)0} = 0$ & $-$ \\
				$|\varphi_{d_3+d_1-4}\rangle,\ |\varphi_{3d_3+d_2+2d_1-12}\rangle$ & $a_{0(d_1-1)} = a_{(d_1-1)0} = 0$ & $-$ \\
				\hline
			\end{tabular}
		\end{center}
		\label{tab:Nondiagonal_elements}
		
		\begin{center}
			\begin{tabular}{c|c|c} 
				
				\multicolumn{3}{c}{\textbf{TABLE 8. Diagonal elements in $M_A^\dagger M_A$.}} \\
				\hline\hline
				
				Pair of states & Diagonal elements & Ranges \\
				\hline
				
				$|\varphi_i\rangle,\ |\varphi_{4d_3+2d_1-11}\rangle$ & $a_{(i+1)(i+1)} = a_{(d_1-1)(d_1-1)} $ & $1 \le i \le d_1-3$ \\
				$|\varphi_{d_1-2}\rangle,\ |\varphi_{4d_3+2d_1-11}\rangle$ & $a_{11} = a_{(d_1-1)(d_1-1)} $ & $-$ \\
				$|\varphi_{3d_3+2d_1-11}\rangle,\ |\varphi_{4d_3+2d_1-11}\rangle$ & $a_{00} = a_{(d_1-2)(d_1-2)}$ & $-$ \\
				\hline
			\end{tabular}
		\end{center}
		\label{tab:Diagonal_elements}
		
		From Tables \textcolor{blue}{7} and \textcolor{blue}{8}, we see that any POVM element $M_A^\dagger M_A$ is proportional to the identity matrix. Therefore, Alice can only perform a trivial measurement. 

        On the other hand, suppose Bob first performs an orthogonality-preserving measurement with a set of POVM elements $\{M_B^\dagger M_B\}$, where each \(M_B^\dagger M_B\) can be written as
        \[
		M_B^\dagger M_B =
		\begin{bmatrix}
			b_{00} & b_{01} & \cdots & b_{0(d_2-1)} \\
			b_{10} & b_{11} & \cdots & b_{1(d_2-1)} \\
			\vdots & \vdots & \ddots & \vdots \\
			b_{(d_2-1)0} & b_{(d_2-1)1} & \cdots & b_{(d_2-1)(d_2-1)}
		\end{bmatrix}_{d_2\times d_2}
		\]
		under the basis $\{|0\rangle,\,|1\rangle,\,\cdots,\,|(d_2-1)\rangle\}$.
		
		Similar to the prior proof, the corresponding results are given in Tables \textcolor{blue}{9} and \textcolor{blue}{10}.
		
		\begin{center}
			\begin{tabular}{c|c|c} 
				
				\multicolumn{3}{c}{\textbf{TABLE 9. Nondiagonal elements in $M_B^\dagger M_B$.}} \\
				\hline\hline
				
				Pair of states & Nondiagonal elements & Ranges \\
				\hline
				
				$|\varphi_{d_3+d_1-4+i}\rangle,\ |\varphi_{d_3+d_1-4+j}\rangle$ & $b_{(d_2-2-i)(d_2-2-j)} = b_{(d_2-2-j)(d_2-2-i)} =0$ & $1 \le i \neq j \le d_2-3$ \\
				$|\varphi_{d_3+d_1-4+i}\rangle,\ |\varphi_{d_3+d_2+d_1-6}\rangle$ & $b_{0(d_2-2-i)} = b_{(d_2-2-i)0} =0$ & $1 \le i \le d_2-3 $ \\
				$|\varphi_{d_3+d_1-4+i}\rangle,\ |\varphi_{i+d_3+d_2+d_1-6}\rangle$ & $b_{0(d_2-2)} = b_{(d_2-2)0} = 0$ & $-$ \\
				$|\varphi_{d_3+d_2+d_1-6}\rangle,\ |\varphi_{i+d_3+d_2+d_1-6}\rangle$ & $b_{(d_2-2)(d_2-2-i)} = b_{(d_2-2-i)(d_2-2)} = 0$ & $1 \le i \le d_2-3$ \\
				$|\varphi_{3d_3+d_2+2d_1-13}\rangle,\ |\varphi_{3d_3+d_1-9+i}\rangle$ & $b_{(d_2-1)(1+i)} = b_{(1+i)(d_2-1)} = 0$ & $1 \le i \le d_1-3$ \\
				$|\varphi_{4d_3+3d_2-14}\rangle,\ |\varphi_{3d_3+2d_1-11}\rangle$ & $b_{(d_2-1)(d_1-1)} = b_{(d_1-1)(d_2-1)} = 0$ & $-$ \\
				$|\varphi_{3d_3+2d_2+d_1-14+i}\rangle,\ |\varphi_{3d_3+2d_2+d_1-14+j}\rangle$ & $b_{(d_1-1+i)(d_1-1+j)} = b_{(d_1-1+j)(d_1-1+i)} =0$ & $1 \le i \neq j \le d_2-d_1$ \\
				$|\varphi_{3d_3+2d_1-11}\rangle,\ |\varphi_{3d_3+d_2+2d_1-13}\rangle$ & $b_{1(d_2-1)} = b_{(d_1-1)1} = 0,\ b_{(d_2-1)1} = b_{1(d_1-1)} = 0$ & $-$ \\
				$|\varphi_{3d_3+d_2+2d_1-13}\rangle,\ |\varphi_{3d_3+d_2+2d_1-12}\rangle$ & $b_{0(d_2-1)} = b_{(d_2-1)0} = 0$ & $-$ \\
				\hline
			\end{tabular}
		\end{center}
		\label{tab:Nondiagonal_elements}
		
		\begin{center}
			\begin{tabular}{c|c|c} 
				
				\multicolumn{3}{c}{\textbf{TABLE 10. Diagonal elements in $M_B^\dagger M_B$.}} \\
				\hline\hline
				
				Pair of states & Diagonal elements & Ranges \\
				\hline
				
				$|\varphi_{d_3+d_1-4+i}\rangle,\ |\varphi_{4d_3+2d_1-11}\rangle$ & $b_{00} = b_{(d_2-2-i)(d_2-2-i)} $ & $1 \le i \le d_2-3$ \\
				$|\varphi_{2d-3+d_2+d_1-8}\rangle,\ |\varphi_{4d_3+2d_1-11}\rangle$ & $b_{00} = b_{(d_2-2)(d_2-2)} $ & $-$ \\
				$|\varphi_{3d_3+d_2+2d_1-13}\rangle,\ |\varphi_{4d_3+2d_1-11}\rangle$ & $b_{11} = b_{(d_2-1)(d_2-1)}$ & $-$ \\
				\hline
			\end{tabular}
		\end{center}
		\label{tab:Diagonal_elements}
		
		From Tables \textcolor{blue}{9} and \textcolor{blue}{10}, we see that any POVM element $M_B^\dagger M_B$ is proportional to the identity matrix. Therefore, Bob can only perform a trivial measurement.
 
        Next, suppose Charlie first performs an orthogonality-preserving measurement with a set of POVM elements $\{M_C^\dagger M_C\}$, where each \(M_C^\dagger M_C\) can be written as
		\[
		M_C^\dagger M_C =
		\begin{bmatrix}
			c_{00} & c_{01} & \cdots & c_{0(d_3-1)} \\
			c_{10} & c_{11} & \cdots & c_{1(d_3-1)} \\
			\vdots & \vdots & \ddots & \vdots \\
			c_{(d_3-1)0} & c_{(d_3-1)1} & \cdots & c_{(d_3-1)(d_3-1)}
		\end{bmatrix}_{d_3\times d_3}
		\]
		under the computed basis \(\{|0\rangle,|1\rangle,\cdots,|(d_3-1)\rangle\}\).
		
		If Charlie performs measurements first with orthogonality-preserving POVM elements \(M_C^\dagger M_C\), and the results are listed in Tables \textcolor{blue}{11} and \textcolor{blue}{12}.
		
		\begin{center}
			\begin{tabular}{c|c|c} 
				
				\multicolumn{3}{c}{\textbf{TABLE 11. Nondiagonal elements in $M_C^\dagger M_C$.}} \\
				\hline\hline
				
				Pair of states & Nondiagonal elements & Ranges \\
				\hline
				
				$|\varphi_{d_3+d_2+d_1-6}\rangle,\ |\varphi_{d_3+2d_2+d_1-9+i}\rangle$ & $c_{(d_2-1)(d_2-1+i)} = c_{(d_2-1+i)(d_2-1)} =0$ & $1 \le i \le d_3-d_2$ \\
				$|\varphi_{2d_3+d_2+d_1-8}\rangle,\ |\varphi_{d_3+2d_2+d_1-9+i}\rangle$ & $c_{1(d_2-1+i)} = c_{(d_2-1+i)1} =0$ & $1 \le i \le d_3-d_2 $ \\
				$|\varphi_{d_3+d_2+d_1-6}\rangle,\ |\varphi_{2d_3+d_2+d_1-8}\rangle$ & $c_{(d_2-1)1} = c_{1(d_3-1)} = 0,\ c_{1(d_2-1)} = c_{(d_3-1)1} = 0 $ & $-$ \\
				$|\varphi_{i+d_3+d_2+d_1-6}\rangle,\ |\varphi_{d_3+2d_2+d_1-9+j}\rangle$ & $c_{(1+i)(d_2-1+j)} = c_{(d_2-1+j)(1+i)} = 0$ & $1 \le i \le d_2-3, 1 \le j \le d_3-d_2 $ \\
				$|\varphi_{d_3+2d_2+d_1-9+i}\rangle,\ |\varphi_{d_3+2d_2+d_1-9+j}\rangle$ & $c_{(d_2-1+i)(d_2-1+j)} = c_{(d_2-1+j)(d_2-1+i)} = 0$ & $1 \le i \neq j \le d_3-d_2$ \\
				$|\varphi_{d_3+d_1-4+i}\rangle,\ |\varphi_{d_3+d_1-4+j}\rangle$ & $c_{(1+j)(1+i)} = c_{(1+i)(1+j)} = 0$ & $1 \le i \neq j \le d_2-3$ \\
				$|\varphi_{2d_3+d_2+d_1-8}\rangle,\ |\varphi_{i+d_3+d_2+d_1-6}\rangle$ & $c_{1(1+i)} = c_{(1+i)1} = 0$ & $1 \le i \le d_2-3$ \\
				$|\varphi_{i+d_3+d_2+d_1-6}\rangle,\ |\varphi_{d_3+d_2+d_1-6}\rangle$ & $c_{(d_2-1)(1+i)} = c_{(1+i)(d_2-1)} = 0$ & $1 \le i \le d_2-3$ \\
				$|\varphi_{3d_3+d_2+2d_1-12}\rangle,\ |\varphi_{d_3+d_2+d_1-6}\rangle$ & $c_{10} = c_{01} = 0$ & $-$ \\
				$|\varphi_{3d_3+d_2+2d_1-12}\rangle,\ |\varphi_{d_3+d_1-4+i}\rangle$ & $c_{0(1+i)} = c_{(1+i)0} =0$ & $1 \le i  \le d_2-3$ \\
				$|\varphi_{3d_3+d_2+2d_1-12}\rangle,\ |\varphi_{3d_3+d_2+2d_1-12+i}\rangle$ & $c_{0(d_2-2+i)} = c_{(d_2-2+i)0} = 0$ & $1 \le i \le d_3-d_2$ \\
				$|\varphi_{3d_3+d_2+2d_1-12}\rangle,\ |\varphi_{2d_3+d_2+d_1-8}\rangle$ & $c_{(d_3-1)0} = c_{0(d_3-1)} = 0$ & $-$ \\
				\hline
			\end{tabular}
		\end{center}
		\label{tab:Nondiagonal_elements}
		
		\begin{center}
			\begin{tabular}{c|c|c} 
				
				\multicolumn{3}{c}{\textbf{TABLE 12. Diagonal elements in $M_C^\dagger M_C$.}} \\
				\hline\hline
				
				Pair of states & Diagonal elements & Ranges \\
				\hline
				
				$|\varphi_{3d_3+d_2+2d_1-12}\rangle,\ |\varphi_{4d_3+2d_1-11}\rangle$ & $c_{00} = c_{11} $ & $-$ \\
				$|\varphi_{d_3+d_1-4+i}\rangle,\ |\varphi_{4d_3+2d_1-11}\rangle$ & $c_{11} = c_{(1+i)(1+i)} $ & $1 \le i \le d_2-3$ \\
				$|\varphi_{d_3+d_2+d_1-6}\rangle,\ |\varphi_{4d_3+2d_1-11}\rangle$ & $c_{11} = c_{(d_2-1)(d_2-1)}$ & $-$ \\
                $|\varphi_{2d_3+d_2+d_1-8}\rangle,\ |\varphi_{4d_3+2d_1-11}\rangle$ & $c_{11} = c_{(d_3-1)(d_3-1)}$ & $-$ \\
				\hline
			\end{tabular}
		\end{center}
		\label{tab:Diagonal_elements}
		 \end{widetext}
		 
		Similar systems of equations for Theorem \hyperref[Theorem2]{2} can also be derived from $|\phi_{2d_3+d_2+d_1-8+i}\rangle$ and $|\phi_{4d_3+2d_1-11}\rangle$:
		\begin{align*}
			\begin{cases}
				\displaystyle
                \sum_{j=0}^{d_{3}-d_{2}-1}\omega^{j}c_{(d_{2}+j)(d_{2}+j)}=0,\\
                \displaystyle
                \sum_{j=0}^{d_{3}-d_{2}-1}\omega^{2j}c_{(d_{2}+j)(d_{2}+j)}=0,\\
				\quad\vdots\\[6pt]
				\displaystyle
				\sum_{j=0}^{d_{3}-d_{2}-1}\omega^{(d_{3}-d_{2}-1)j}c_{(d_{2}+j)(d_{2}+j)}=0,
			\end{cases}
			\tag{3}\label{eq:sys3}
		\end{align*}
		where  $\omega = e^{\frac{2\pi \sqrt{-1}}{d_3-d_2}}$.
		
		By Lemma \hyperref[lemma1]{1}, this system of Eq. \eqref{eq:sys3} has a unique solution, namely
		$c_{d_2d_2} = c_{(d_2+1)(d_2+1)} = \cdots = c_{(d_3-1)(d_3-1)}$. Combined with Table \textcolor{blue}{11} and Table \textcolor{blue}{12}, we see that any POVM element $M_C^\dagger M_C$ is proportional to the identity matrix. Therefore, Charlie can only perform a trivial measurement. In summary, we have successfully proved that $S_3$ cannot be distinguished by LOCC.
		
		Now, we prove that $S_3 - \{|\varphi_{4d_3+2d_1-11}\rangle\}$ is distinguishable under LOCC.
		As the first step, Alice performs a measurement with the operators: $\{A_0 = |0\rangle_A\langle 0|, \,A_1 = I_A-A_0\}$ .
		
		(1) If Alice’s measurement outcome corresponds to $A_0 = |0\rangle_A \langle 0|$, the state measured must be one of
		$\{|\varphi_{d_3+d_1-4+i}\rangle,$ $|\varphi_{d_3+d_2+d_1-6}\rangle,$ $|\varphi_{i+d_3+d_2+d_1-6}\rangle,$ $|\varphi_{d_3+2d_2+d_1-9+i}\rangle,$ 
$ |\varphi_{2d_3+d_2+d_1-8}\rangle,$ $|\varphi_{2d_3+d_2+d_1-8+i}\rangle,$ $|\varphi_{3d_3+d_1-9+i}\rangle,$ $|\varphi_{3d_3+2d_1-11}\rangle,$ $|\varphi_{3d_3+d_2+2d_1-13}\rangle,$
$ |\varphi_{3d_3+d_2+2d_1-12}\rangle,$ $|\varphi_{3d_3+d_2+2d_1-12+i}\rangle\}$, and it will collapse to one of the following forms:
		\begin{align}
			|\varphi_{d_3+d_1-4+i}\rangle \to &|\varphi_{d_3+d_1-4+i}\rangle, \nonumber\\
			& 1 \le i \le d_2-3; \nonumber\\
			|\varphi_{d_3+d_2+d_1-6}\rangle \to &|\varphi_{d_3+d_2+d_1-6}\rangle; \nonumber\\
			|\varphi_{i+d_3+d_2+d_1-6}\rangle \to&|\varphi_{i+d_3+d_2+d_1-6}\rangle, \nonumber\\
			& 1 \le i\le d_2-3; \nonumber\\
			|\varphi_{d_3+2d_2+d_1-9+i}\rangle \to &|\varphi_{d_3+2d_2+d_1-9+i}\rangle ,\nonumber \\
			& 1 \le i \le d_3-d_2;\nonumber \\
			|\varphi_{2d_3+d_2+d_1-8}\rangle \to &|\varphi_{2d_3+d_2+d_1-8}\rangle;\nonumber \\
			|\varphi_{2d_3+d_2+d_1-8+i}\rangle \to &|\varphi_{2d_3+d_2+d_1-8+i}\rangle,\nonumber \\
			& 1 \le i \le d_3-d_2-1 ;\nonumber \\
			|\varphi_{3d_3+d_1-9+i}\rangle \to &|0(1+i)0\rangle_{ABC},\nonumber \\
			& 1\le i \le d_1-3 ;\nonumber \\
			|\varphi_{3d_3+2d_1-11}\rangle \to &|010\rangle_{ABC};\nonumber \\
			|\varphi_{3d_3+d_2+2d_1-13}\rangle \to &|0(d_2-1)0\rangle_{ABC};\nonumber \\
			|\varphi_{3d_3+d_2+2d_1-12}\rangle \to &|000\rangle_{ABC};\nonumber \\
			|\varphi_{3d_3+d_2+2d_1-12+i}\rangle \to &|\varphi_{3d_3+d_2+2d_1-12+i}\rangle , \nonumber \\
			 & 1 \le i \le d_3-d_2.\nonumber 
		\end{align}
		
		Bob then implements a measurement with the measurement operators: $B_0 = |0\rangle_B\langle 0|,$ $B_1 = |1\rangle_B\langle 1|,$ $B_2 = |(d_2-4)\rangle_B\langle (d_2-4)|,$ $B_3 = |(d_2-3)\rangle_B\langle( d_2-3)|,$ $B_4 = |(d_2-2)\rangle_B\langle (d_2-2)|,$ $B_5 = |(d_2-1)\rangle_B\langle (d_2-1)|,$ $B_6 =I_B-B_0-B_1-B_2-B_3-B_4-B_5 $.
		
		\textcircled{1} If Bob’s measurement outcome corresponds to $B_0$, the measured state will collapse to one of the following forms:
		\begin{align}
			|\varphi_{d_3+d_1-4+i}\rangle \to &|00(1+i)\rangle_{ABC}, \nonumber\\
			&1 \le i \le d_2-3; \nonumber\\
			|\varphi_{d_3+d_2+d_1-6}\rangle \to &|001\rangle_{ABC}; \nonumber\\
			|\varphi_{2d_3+d_2+d_1-8}\rangle \to &|00(d_3-1)\rangle_{ABC};\nonumber \\
			|\varphi_{3d_3+d_2+2d_1-12}\rangle \to &|000\rangle_{ABC};\nonumber \\
			|\varphi_{3d_3+d_2+2d_1-12+i}\rangle \to &|0\rangle_A |0\rangle_B|(d_2-2+i)\rangle_C,\nonumber \\
			&1 \le i \le d_3-d_2.\nonumber 
		\end{align}
		
		All collapsed states are mutually orthogonal on Charlie’s subsystem. So, they are distinguishable via LOCC operations.
		
		\textcircled{2} If Bob’s measurement outcome corresponds to $B_1$, the measured state will collapse to one of the following forms:
		\begin{align}
			|\varphi_{d_3+d_1+d_2-7}\rangle &\to |011\rangle_{ABC};  \nonumber\\
			|\varphi_{d_3+2d_2+d_1-9}\rangle &\to |01(d_2-1)\rangle_{ABC};  \nonumber\\
			|\varphi_{3d_3+2d_1-11}\rangle &\to |010\rangle_{ABC}.\nonumber 
		\end{align}
		
		All the collapsed states are mutually orthogonal on Charlie's subsystem, so these states can be distinguished by LOCC.
		
		\textcircled{3} If Bob’s measurement outcome corresponds to $B_2$, the measured state will collapse to one of the following forms:
		\begin{align}
			|\varphi_{d_3+d_1-2}\rangle \to &|0(d_2-4)1\rangle_{ABC};  \nonumber\\
			|\varphi_{d_3+d_2+d_1-4}\rangle \to &|0(d_2-4)(d_2-1)\rangle_{ABC};  \nonumber\\
			|\varphi_{2d_3+d_2+d_1-8+i}\rangle \to &|\varphi_{2d_3+d_2+d_1-8+i}\rangle, \nonumber  \\
			& 1 \le i \le d_3-d_2-1. \nonumber  
		\end{align}
		
		All collapsed states are mutually orthogonal on Charlie’s subsystem. So, they are distinguishable via LOCC operations.
		
		\textcircled{4} If Bob’s measurement outcome corresponds to $B_3$, the measured state must be one of:
		\begin{align}
			|\varphi_{d_3+d_1-3}\rangle \to &|0(d_2-3)1\rangle_{ABC};  \nonumber\\
			|\varphi_{d_3+d_2+d_1-5}\rangle \to &|0(d_2-3)(d_2-1)\rangle_{ABC};  \nonumber\\
			|\varphi_{d_3+2d_2+d_1-9+i}\rangle \to &|0\rangle_A |(d_2-3)\rangle_B|(d_2-1+i)\rangle_C, \nonumber \\
			&1\le i \le d_3-d_2. \nonumber 
		\end{align}
		
		All the collapsed states are mutually orthogonal on Charlie's subsystem, so these states can be distinguished by LOCC.
		
		\textcircled{5} If Bob’s measurement outcome corresponds to $B_4$, the measured state will collapse to one of the following forms:
		\begin{align}
			|\varphi_{d_3+d_2+d_1-6}\rangle \to &|0(d_2-2)(d_2-1)\rangle_{ABC};  \nonumber\\
			|\varphi_{i+d_3+d_2+d_1-6}\rangle \to &|0(d_2-2)(1+i)\rangle_{ABC},  \nonumber\\
			&1 \le i \le d_2-3;  \nonumber\\
			|\varphi_{d_3+2d_2+d_1-9+i}\rangle \to  &|0\rangle_A |(d_2-2)\rangle_B|(d_2-1+i)\rangle_C, \nonumber \\
			&1 \le i \le d_3-d_2; \nonumber \\
			|\varphi_{2d_3+d_2+d_1-8}\rangle &\to |0(d_2-2)1\rangle_{ABC}.  \nonumber
		\end{align}
		
		All collapsed states are mutually orthogonal on Charlie’s subsystem; as a result, so these states can be distinguished by LOCC.
		
		\textcircled{6} If Bob’s measurement outcome corresponds to $B_5$, the measured state will collapse to one of the following forms:
		\begin{align}
			|\varphi_{3d_3+d_2+2d_1-13}\rangle \to &|0(d_2-1)0\rangle_{ABC};\nonumber \\
			|\varphi_{3d_3+d_2+2d_1-12+i}\rangle \to &|0(d_2-1)(d_2-2+i)\rangle_{ABC} ,\nonumber\\
			&1 \le i \le d_3-d_2.\nonumber
		\end{align}
		
		All the collapsed states are mutually orthogonal on Charlie's subsystem, so these states can be distinguished by LOCC.
		
		\textcircled{7} If Bob’s measurement outcome corresponds to $B_6$, the measured state must be one of the following forms:
		\begin{align}
			|\varphi_{d_3+d_1-4+i}\rangle \to &|0(d_2-2-i)1\rangle_{ABC}, \nonumber \\
			 & 3 \le i \le d_2-4; \nonumber \\
			|\varphi_{i+d_3+d_2+d_1-6}\rangle \to &|0(d_2-2-i)(d_2-1)\rangle_{ABC} ,\nonumber\\
			 & 3 \le i \le d_2-4;\nonumber\\
			|\varphi_{3d_3+d_1-9+i}\rangle \to &|0(1+i)0\rangle_{ABC} ,\nonumber\\
			 & 1 \le i \le d_1-3.\nonumber
		\end{align}
		
		All collapsed states are mutually orthogonal on Charlie’s subsystem. So, they are distinguishable via LOCC operations.
		
		(2) If Alice’s measurement outcome corresponds to $A_1=I_A-A_0$, the state measured must be one of the following forms:
		$\{|\varphi_{i}\rangle,$ $|\varphi_{d_1-2}\rangle,$ $|\varphi_{i+d_1-2}\rangle,$ $|\varphi_{2d_1-5+i}\rangle,$ $ |\varphi_{d_3+d_1-4}\rangle,$ $|\varphi_{3d_3+d_1-9+i}\rangle,$ $ |\varphi_{3d_3+2d_1-11}\rangle,$ $ |\varphi_{i+3d_3+2d_1-11}\rangle,$ $|\varphi_{3d_3+2d_2+d_1-14+i}\rangle,$ $ |\varphi_{3d_3+d_2+2d_1-13}\rangle,$ 
$|\varphi_{3d_3+d_2+2d_1-12}\rangle\}$, it will collapse to one of the following forms:
		\begin{align}
			|\varphi_i\rangle \to &|\varphi_i\rangle,\nonumber  \\
			 & 1 \le i \le d_1-3;\nonumber  \\
			|\varphi_{d_1-2}\rangle \to &|\varphi_{d_1-2}\rangle; \nonumber \\
			|\varphi_{i+d_1-2}\rangle \to & |\varphi_{i+d_1-2}\rangle,\nonumber  \\
			 &1 \le i \le d_1-3;\nonumber  \\
			|\varphi_{2d_1-5+i}\rangle \to &|\varphi_{2d_1-5+i}\rangle, \nonumber \\
			 &1 \le i \le d_3-d_1; \nonumber \\
			|\varphi_{d_3+d_1-4}\rangle \to &|\varphi_{d_3+d_1-4}\rangle;\nonumber \\
			|\varphi_{3d_3+d_1-9+i}\rangle \to &|(d_1-2-i)10\rangle_{ABC} ,\nonumber  \\
			& 1 \le i \le d_1-3;\nonumber  \\
			|\varphi_{3d_3+2d_1-11}\rangle \to &|(d_1-2)(d_1-1)0\rangle_{ABC}; \nonumber \\
			|\varphi_{i+3d_3+2d_1-11}\rangle \to &|\varphi_{i+3d_3+2d_1-11}\rangle,\nonumber  \\
			 & 1 \le i \le d_1-3;\nonumber  \\
			|\varphi_{3d_3+2d_2+d_1-14+i}\rangle \to &|\varphi_{3d_3+2d_2+d_1-14+i}\rangle, \nonumber \\
			 &1 \le i \le d_2-d_1; \nonumber \\
			|\varphi_{3d_3+d_2+2d_1-13}\rangle \to &|(d_1-2)10\rangle_{ABC};\nonumber \\
			|\varphi_{3d_3+d_2+2d_1-12}\rangle \to &|111\rangle_{ABC}. \nonumber
		\end{align}
		
		Bob then performs a measurement with the POVM operators: $B_0 = |0\rangle_B\langle 0|,\, B_1 = |1\rangle_B\langle 1| ,\, B_2 = I_B-B_0-B_1$.
		
		\textcircled{1} If Bob’s measurement outcome corresponds to $B_0$, the measured state will collapse to one of the following forms:
		\begin{align}
			|\varphi_i\rangle &\to |\varphi_i\rangle,& 1 &\le i \le d_1-3; \nonumber  \\
			|\varphi_{d_1-2}\rangle &\to |\varphi_{d_1-2}\rangle; \nonumber \\
			|\varphi_{i+d_1-2}\rangle &\to |\varphi_{i+d_1-2}\rangle,& 1 &\le i \le d_1-3;\nonumber  \\
			|\varphi_{2d_1-5+i}\rangle &\to |\varphi_{2d_1-5+i}\rangle, &1 &\le i \le d_3-d_1; \nonumber \\
			|\varphi_{d_3+d_1-4}\rangle &\to |\varphi_{d_3+d_1-4}\rangle .\nonumber 
		\end{align}
		
		Since collapsed states have zero components on Bob's subsystem, we can omit Bob's subsystem. The simplified states are equivalent to the constructions for Alice and Charlie subsystems in Theorem \hyperref[Theorem2]{2}. Therefore, these collapsed states can be distinguished by LOCC.
		
		\textcircled{2} If Bob’s measurement outcome corresponds to $B_1$, the measured state will collapse to one of the following forms:
		\begin{align}
			|\varphi_{3d_3+d_1-9+i}\rangle \to &|(d_1-2-i)10\rangle_{ABC} ,\nonumber  \\ 
			&1 \le i \le d_1-3;\nonumber  \\ 
			|\varphi_{3d_3+d_2+2d_1-13}\rangle \to &|(d_1-2)10\rangle_{ABC};\nonumber \\
			|\varphi_{3d_3+d_2+2d_1-12}\rangle \to &|111\rangle_{ABC}. \nonumber
		\end{align}		
		These collapsed states can be perfectly distinguished by LOCC by Alice and Charlie.
		
		\textcircled{3} If Bob’s measurement outcome corresponds to $B_2$, the measured state will collapse to one of the following forms:
		\begin{align}
			|\varphi_{3d_3+2d_1-11}\rangle \to &|(d_1-2)(d_1-1)0\rangle_{ABC} ; \nonumber \\
			|\varphi_{i+3d_3+2d_1-11}\rangle \to &|\varphi_{i+3d_3+2d_1-11}\rangle,\nonumber  \\
			 & 1 \le i \le d_1-3;\nonumber  \\
			|\varphi_{3d_3+2d_2+d_1-14+i}\rangle \to &|\varphi_{3d_3+2d_2+d_1-14+i}\rangle, \nonumber \\
			& 1 \le i \le d_2-d_1.\nonumber 
		\end{align}
		
		The remaining states have zero components on Charlie's subsystem, so we can omit it. The simplified states conform to the constructions in Theorem \hyperref[Theorem2]{2}. Hence, these collapsed states can also be distinguished by LOCC.
		
	 We have fully proved that $S_3 - \{|\varphi_{4d_3+2d_1-11}\rangle\}$ is distinguishable by LOCC. Therefore, $S_3$ possesses minimal nonlocality. This completes the proof.
\end{proof}
	
	 To illustrate the construction of the tripartite structure clearly, we take the $ \mathbb C^{5} \otimes \mathbb C^{7}\otimes \mathbb C^{9}$ quantum system as an example.
		\begin{figure}[htbp]
	\vspace{2mm}
	\centering
	\begin{tikzpicture}[
		scale=0.77,
		line width=0.5mm,
		x={(1cm, 0cm)},
		y={(-0.4cm, 0.4cm)},
		z={(0cm, 1cm)},
		line cap=round, line join=round
		]
		
		\node at (2.5, 0, -1)    {\textbf{Alice}};
		\node at (-1.5, 3.5, 0)  {\textbf{Bob}};
		\node at (8, 3.5, 2)   {\footnotesize \textbf{Charlie}};
		
		\node at (0.6, 0, -0.5)  {\footnotesize $|0\rangle$};
		\node at (1.6, 0, -0.5)  {\footnotesize $|1\rangle$};
		\node at (2.6, 0, -0.5)  {\footnotesize $|2\rangle$};
		\node at (3.6, 0, -0.5)  {\footnotesize $|3\rangle$};
		\node at (4.6, 0, -0.5)  {\footnotesize $|4\rangle$};
		
		\node at (0, 1, -0.5)    {\footnotesize $|0\rangle$};
		\node at (0, 2, -0.5)    {\footnotesize $|1\rangle$};
		\node at (0, 3.2, -0.5)  {\footnotesize $|2\rangle$};
		\node at (0, 4.2, -0.5)  {\footnotesize $|3\rangle$};
		\node at (0, 5.2, -0.5)  {\footnotesize $|4\rangle$};
		\node at (-0.2, 6.2, -0.5){\footnotesize $|5\rangle$};
		\node at (-0.2, 7.2, -0.5){\footnotesize $|6\rangle$};
		
		\node at (7, 4, -1.2)     {\footnotesize $|0\rangle$};
		\node at (7, 4, -0.1)     {\footnotesize $|1\rangle$};
		\node at (7, 4, 0.9)      {\footnotesize $|2\rangle$};
		\node at (7, 4, 1.9)    {\footnotesize $|3\rangle$};
		\node at (7, 4, 2.9)    {\footnotesize $|4\rangle$};
		\node at (7, 4, 3.9)    {\footnotesize $|5\rangle$};
		\node at (7, 4, 4.9)    {\footnotesize $|6\rangle$};
		\node at (7, 4, 5.9)    {\footnotesize $|7\rangle$};
		\node at (7, 4, 6.9)    {\footnotesize $|8\rangle$};
		
		\fill[teal!25]   (0,4,9) -- (1,4,9) -- (1,5,9) -- (0,5,9) -- cycle;
		\fill[teal!25]   (0,5,9) -- (1,5,9) -- (1,6,9) -- (0,6,9) -- cycle;
		\fill[violet!25] (0,0,9) -- (1,0,9) -- (1,1,9) -- (0,1,9) -- cycle;
		
		\fill[gray!15]   (1,0,0) rectangle (2,0,1);
		\node at (1.5,0,0.5) {\scriptsize 9};
		\fill[gray!15]   (2,0,0) rectangle (3,0,1);
		\node at (2.5,0,0.5) {\scriptsize 9};
		\fill[pink!15]   (4,0,0) rectangle (5,0,1);
		\node at (4.5,0,0.5) {\scriptsize 10};
		
		\fill[cyan!15]  (0,0,1) rectangle (1,0,2);
		\fill[brown!15]  (1,0,1) rectangle (2,0,2);
		\node at (1.5,0,1.5) {\scriptsize 8};
		\fill[brown!15]  (2,0,1) rectangle (3,0,2);
		\node at (2.5,0,1.5) {\scriptsize 8};
		
		\fill[red!25] (0,0,2) rectangle (1,0,3);
		\fill[purple!15] (1,0,2) rectangle (2,0,3);
		\node at (1.5,0,2.5) {\scriptsize 7};
		\fill[purple!15] (2,0,2) rectangle (3,0,3);
		\node at (2.5,0,2.5) {\scriptsize 7};
		
		\fill[green!15] (0,0,3) rectangle (1,0,4);
		\fill[yellow!15] (1,0,3) rectangle (2,0,4);
		\node at (1.5,0,3.5) {\scriptsize 6};
		\fill[yellow!15] (2,0,3) rectangle (3,0,4);
		\node at (2.5,0,3.5) {\scriptsize 6};
		
		\fill[blue!15]   (0,0,4) rectangle (1,0,5);
		\fill[blue!15]   (1,0,4) rectangle (2,0,5);
		\node at (1.5,0,4.5) {\scriptsize 3};
		\fill[orange!15] (2,0,4) rectangle (3,0,5);
		\node at (2.5,0,4.5) {\scriptsize 4};
		\fill[cyan!15]   (3,0,4) rectangle (4,0,5);
		\node at (3.5,0,4.5) {\scriptsize 5};
		
		\fill[orange!15]   (0,0,5) rectangle (1,0,6);
		\fill[cyan!15]   (1,0,5) rectangle (2,0,6);
		\node at (1.5,0,5.5) {\scriptsize 5};
		\fill[green!15]  (4,0,5) rectangle (5,0,6);
		\node at (4.5,0,5.5) {\scriptsize 2};
		
		\fill[blue!35] (0,0,6) rectangle (1,0,7);
		\fill[orange!15] (1,0,6) rectangle (2,0,7);
		\node at (1.5,0,6.5) {\scriptsize 4};
		\fill[red!25]    (4,0,6) rectangle (5,0,7);
		\node at (4.5,0,6.5) {\scriptsize 1};
		
		\fill[red!55]   (0,0,7) rectangle (1,0,8);
		\fill[pink!15]   (1,0,7) rectangle (2,0,8);
		\node at (1.5,0,7.5) {\scriptsize 10};
		\fill[red!25]    (2,0,7) rectangle (3,0,8);
		\node at (2.5,0,7.5) {\scriptsize 1};
		\fill[green!15]  (3,0,7) rectangle (4,0,8);
		\node at (3.5,0,7.5) {\scriptsize 2};
		\fill[blue!15]   (4,0,7) rectangle (5,0,8);
		\node at (4.5,0,7.5) {\scriptsize 3};
		
		\fill[violet!25]   (0,0,8) rectangle (1,0,9);
		
		
		\fill[teal!25]   (0,5,8) -- (0,6,8) -- (0,6,9) -- (0,5,9) -- cycle; \node at (0,5.5,8.5){\scriptsize21};
		\fill[teal!25]   (0,4,8) -- (0,5,8) -- (0,5,9) -- (0,4,9) -- cycle; \node at (0,4.5,8.5){\scriptsize21};
		\fill[violet!25] (0,0,8) -- (0,1,8) -- (0,1,9) -- (0,0,9) -- cycle; \node at (0,0.5,8.5){\scriptsize22};
		
		\fill[red!55]    (0,6,7) -- (0,7,7) -- (0,7,8) -- (0,6,8) -- cycle; \node at (0,6.5,7.5){\scriptsize34};
		\fill[pink!15]   (0,5,7) -- (0,6,7) -- (0,6,8) -- (0,5,8) -- cycle; \node at (0,5.5,7.5){\scriptsize20};
		\fill[pink!15]   (0,4,7) -- (0,5,7) -- (0,5,8) -- (0,4,8) -- cycle; \node at (0,4.5,7.5){\scriptsize20};
		\fill[red!55]    (0,0,7) -- (0,1,7) -- (0,1,8) -- (0,0,8) -- cycle; \node at (0,0.5,7.5){\scriptsize34};
		
		\fill[blue!35]   (0,6,6) -- (0,7,6) -- (0,7,7) -- (0,6,7) -- cycle; \node at (0,6.5,6.5){\scriptsize33};
		\fill[cyan!15]   (0,5,6) -- (0,6,6) -- (0,6,7) -- (0,5,7) -- cycle; \node at (0,5.5,6.5){\scriptsize15};
		\fill[yellow!15] (0,4,6) -- (0,5,6) -- (0,5,7) -- (0,4,7) -- cycle; \node at (0,4.5,6.5){\scriptsize16};
		\fill[purple!15] (0,3,6) -- (0,4,6) -- (0,4,7) -- (0,3,7) -- cycle; \node at (0,3.5,6.5){\scriptsize17};
		\fill[brown!15]  (0,2,6) -- (0,3,6) -- (0,3,7) -- (0,2,7) -- cycle; \node at (0,2.5,6.5){\scriptsize18};
		\fill[gray!15]   (0,1,6) -- (0,2,6) -- (0,2,7) -- (0,1,7) -- cycle; \node at (0,1.5,6.5){\scriptsize19};
		\fill[blue!35]   (0,0,6) -- (0,1,6) -- (0,1,7) -- (0,0,7) -- cycle; \node at (0,0.5,6.5){\scriptsize33};
		
		\fill[gray!15]   (0,5,5) -- (0,6,5) -- (0,6,6) -- (0,5,6) -- cycle; \node at (0,5.5,5.5){\scriptsize19};
		\fill[orange!15] (0,0,5) -- (0,1,5) -- (0,1,6) -- (0,0,6) -- cycle; \node at (0,0.5,5.5){\scriptsize14};
		
		\fill[brown!15]  (0,5,4) -- (0,6,4) -- (0,6,5) -- (0,5,5) -- cycle; \node at (0,5.5,4.5){\scriptsize18};
		\fill[blue!15]   (0,0,4) -- (0,1,4) -- (0,1,5) -- (0,0,5) -- cycle; \node at (0,0.5,4.5){\scriptsize13};
		
		\fill[purple!15] (0,5,3) -- (0,6,3) -- (0,6,4) -- (0,5,4) -- cycle; \node at (0,5.5,3.5){\scriptsize17};
		\fill[green!15]  (0,0,3) -- (0,1,3) -- (0,1,4) -- (0,0,4) -- cycle; \node at (0,0.5,3.5){\scriptsize12};
		
		\fill[yellow!15] (0,5,2) -- (0,6,2) -- (0,6,3) -- (0,5,3) -- cycle; \node at (0,5.5,2.5){\scriptsize16};
		\fill[red!25]    (0,0,2) -- (0,1,2) -- (0,1,3) -- (0,0,3) -- cycle; \node at (0,0.5,2.5){\scriptsize11};
		
		\fill[violet!25] (0,5,1) -- (0,6,1) -- (0,6,2) -- (0,5,2) -- cycle; \node at (0,5.5,1.5){\scriptsize22};
		\fill[red!25]    (0,4,1) -- (0,5,1) -- (0,5,2) -- (0,4,2) -- cycle; \node at (0,4.5,1.5){\scriptsize11};
		\fill[green!15]  (0,3,1) -- (0,4,1) -- (0,4,2) -- (0,3,2) -- cycle; \node at (0,3.5,1.5){\scriptsize12};
		\fill[blue!15]   (0,2,1) -- (0,3,1) -- (0,3,2) -- (0,2,2) -- cycle; \node at (0,2.5,1.5){\scriptsize13};
		\fill[orange!15] (0,1,1) -- (0,2,1) -- (0,2,2) -- (0,1,2) -- cycle; \node at (0,1.5,1.5){\scriptsize14};
		\fill[cyan!15]   (0,0,1) -- (0,1,1) -- (0,1,2) -- (0,0,2) -- cycle; \node at (0,0.5,1.5){\scriptsize15};
		
		\fill[brown!15]  (0,6,0) -- (0,7,0) -- (0,7,1) -- (0,6,1) -- cycle; 
		\fill[green!15]  (0,3,0) -- (0,4,0) -- (0,4,1) -- (0,3,1) -- cycle;
		\fill[red!25]    (0,2,0) -- (0,3,0) -- (0,3,1) -- (0,2,1) -- cycle; 
		\fill[blue!15]   (0,1,0) -- (0,2,0) -- (0,2,1) -- (0,1,1) -- cycle;
		
		\foreach \i in {0,...,5} {
			\draw (\i, 0, 0) -- (\i, 0, 9);
		}
		\foreach \k in {0,...,9} {
			\draw (0, 0, \k) -- (5, 0, \k);
		}
		
		\foreach \j in {0,...,7} {
			\draw (0, \j, 0) -- (0, \j, 9);
		}
		\foreach \k in {0,...,9} {
			\draw (0, 0, \k) -- (0, 7, \k);
		}
		
		\foreach \i in {0,...,5} {
			\draw (\i, 0, 9) -- (\i, 7, 9);
		}
		\foreach \j in {0,...,7} {
			\draw (0, \j, 9) -- (5, \j, 9);
		}
		
		\draw[thick]
		(0,0,0) -- (5,0,0) -- (5,0,9) -- (0,0,9) -- cycle
		(0,0,0) -- (0,7,0) -- (0,7,9) -- (0,0,9) -- cycle
		(0,0,9) -- (5,0,9) -- (5,7,9) -- (0,7,9) -- cycle;
		
		\begin{scope}[shift={(-2,-10.5,0)}]
			
			\fill[brown!15]  (0,6,0) -- (1,6,0) -- (1,7,0) -- (0,7,0) -- cycle;
			\node at (0.5,6.5,0) {\scriptsize 31};
			\fill[purple!15] (2,6,0) -- (3,6,0) -- (3,7,0) -- (2,7,0) -- cycle;
			\node at (2.5,6.5,0) {\scriptsize 30};
			\fill[purple!15] (3,6,0) -- (4,6,0) -- (4,7,0) -- (3,7,0) -- cycle;
			\node at (3.5,6.5,0) {\scriptsize 30};
			
			\fill[yellow!15] (2,5,0) -- (3,5,0) -- (3,6,0) -- (2,6,0) -- cycle;
			\node at (2.5,5.5,0) {\scriptsize 29};
			\fill[yellow!15] (3,5,0) -- (4,5,0) -- (4,6,0) -- (3,6,0) -- cycle;
			\node at (3.5,5.5,0) {\scriptsize 29};
			
			\fill[cyan!15]   (1,4,0) -- (2,4,0) -- (2,5,0) -- (1,5,0) -- cycle;
			\node at (1.5,4.5,0) {\scriptsize 28};
			\fill[orange!15] (2,4,0) -- (3,4,0) -- (3,5,0) -- (2,5,0) -- cycle;
			\node at (2.5,4.5,0) {\scriptsize 27};
			\fill[blue!15]   (3,4,0) -- (4,4,0) -- (4,5,0) -- (3,5,0) -- cycle;
			\node at (3.5,4.5,0) {\scriptsize 26};
			
			\fill[green!15]  (0,3,0) -- (1,3,0) -- (1,4,0) -- (0,4,0) -- cycle;
			\node at (0.5,3.5,0) {\scriptsize 25};
			\fill[cyan!15]   (3,3,0) -- (4,3,0) -- (4,4,0) -- (3,4,0) -- cycle;
			\node at (3.5,3.5,0) {\scriptsize 28};
			
			\fill[red!25]    (0,2,0) -- (1,2,0) -- (1,3,0) -- (0,3,0) -- cycle;
			\node at (0.5,2.5,0) {\scriptsize 24};
			\fill[orange!15] (3,2,0) -- (4,2,0) -- (4,3,0) -- (3,3,0) -- cycle;
			\node at (3.5,2.5,0) {\scriptsize 27};
			
			\fill[blue!15]   (0,1,0) -- (1,1,0) -- (1,2,0) -- (0,2,0) -- cycle;
			\node at (0.5,1.5,0) {\scriptsize 26};
			\fill[green!15]  (1,1,0) -- (2,1,0) -- (2,2,0) -- (1,2,0) -- cycle;
			\node at (1.5,1.5,0) {\scriptsize 25};
			\fill[red!25]    (2,1,0) -- (3,1,0) -- (3,2,0) -- (2,2,0) -- cycle;
			\node at (2.5,1.5,0) {\scriptsize 24};
			\fill[brown!15]  (3,1,0) -- (4,1,0) -- (4,2,0) -- (3,2,0) -- cycle;
			\node at (3.5,1.5,0) {\scriptsize 31};
			
			\fill[gray!15]   (1,0,0) -- (2,0,0) -- (2,1,0) -- (1,1,0) -- cycle;
			\fill[gray!15]   (2,0,0) -- (3,0,0) -- (3,1,0) -- (2,1,0) -- cycle;
			\fill[pink!15]   (4,0,0) -- (5,0,0) -- (5,1,0) -- (4,1,0) -- cycle;
			
			\foreach \i in {0,...,5} \draw (\i,0,0) -- (\i,7,0);
			\foreach \j in {0,...,7} \draw (0,\j,0) -- (5,\j,0);
		\end{scope}
	\end{tikzpicture}
	\mycaption{The structure of the set $S_4$ in $ \mathbb C^{5} \otimes \mathbb C^{7}\otimes \mathbb C^{9}$ space.}
\end{figure}

	When $d_1=5$, $d_2=7$, $d_3=9$ in Theorem \hyperref[Theorem3]{3}, the following 35 orthogonal quantum states (as shown in Fig. \textcolor{blue}{3}) possess minimal nonlocality:
	
	\begin{align*}
		|\varphi_1\rangle &= |207\rangle_{ABC} - |406\rangle_{ABC}, \nonumber  \\
		|\varphi_2\rangle &= |307\rangle_{ABC} - |405\rangle_{ABC}, \nonumber  \\
		|\varphi_3\rangle &= |407\rangle_{ABC} - |104\rangle_{ABC}, \nonumber  \\
		|\varphi_4\rangle &= |106\rangle_{ABC} - |204\rangle_{ABC}, \nonumber  \\
		|\varphi_5\rangle &= |105\rangle_{ABC} - |304\rangle_{ABC}, \nonumber  \\
		|\varphi_6\rangle &= |1-2\rangle_{A}|0\rangle_{B}|3\rangle_{C}, \nonumber  \\
   		|\varphi_7\rangle &= |1-2\rangle_{A}|0\rangle_{B}|2\rangle_{C}, \nonumber  \\
  		|\varphi_8\rangle &= |1-2\rangle_{A}|0\rangle_{B}|1\rangle_{C}, \nonumber  \\
 		|\varphi_9\rangle &= |1-2\rangle_{A}|0\rangle_{B}|0\rangle_{C}, \nonumber  \\
       	|\varphi_{10}\rangle &= |107\rangle_{ABC} - |400\rangle_{ABC}, \nonumber  \\
    	|\varphi_{11}\rangle &= |041\rangle_{ABC} - |002\rangle_{ABC}, \nonumber  \\
		|\varphi_{12}\rangle &= |031\rangle_{ABC} - |003\rangle_{ABC}, \nonumber  \\
    	|\varphi_{13}\rangle &= |021\rangle_{ABC} - |004\rangle_{ABC}, \nonumber  \\
		|\varphi_{14}\rangle &= |011\rangle_{ABC} - |005\rangle_{ABC}, \nonumber  \\
		|\varphi_{15}\rangle &= |001\rangle_{ABC} - |056\rangle_{ABC}, \nonumber  \\
		|\varphi_{16}\rangle &= |052\rangle_{ABC} - |046\rangle_{ABC}, \nonumber  \\
		|\varphi_{17}\rangle &= |053\rangle_{ABC} - |036\rangle_{ABC}, \nonumber  \\
		|\varphi_{18}\rangle &= |054\rangle_{ABC} - |026\rangle_{ABC}, \nonumber  \\
    	|\varphi_{19}\rangle &= |055\rangle_{ABC} - |016\rangle_{ABC}, \nonumber  \\
		|\varphi_{20}\rangle &= |0\rangle_{A}|5-4\rangle_{B}|7\rangle_{C}, \nonumber  \\
		|\varphi_{21}\rangle &= |0\rangle_{A}|5-4\rangle_{B}|8\rangle_{C}, \nonumber  \\
		|\varphi_{22}\rangle &= |051\rangle_{ABC} - |008\rangle_{ABC}, \nonumber  \\
		|\varphi_{23}\rangle &= |0\rangle_A|3\rangle_B (\left|7\rangle + \omega|8\right\rangle)_C, \nonumber  \\
        |\varphi_{24}\rangle &= |210\rangle_{ABC} - |020\rangle_{ABC}, \nonumber  \\
		|\varphi_{25}\rangle &= |110\rangle_{ABC} - |030\rangle_{ABC}, \nonumber  \\
 		|\varphi_{26}\rangle &= |010\rangle_{ABC} - |340\rangle_{ABC}, \nonumber  \\
      	|\varphi_{27}\rangle &= |320\rangle_{ABC} - |240\rangle_{ABC}, \nonumber  \\
       	|\varphi_{28}\rangle &= |330\rangle_{ABC} - |140\rangle_{ABC}, \nonumber  \\
     	|\varphi_{29}\rangle &= |2-3\rangle_{A}|5\rangle_{B}|0\rangle_{C}, \nonumber  \\
       	|\varphi_{30}\rangle &= |2-3\rangle_{A}|6\rangle_{B}|0\rangle_{C}, \nonumber  \\
       	|\varphi_{31}\rangle &= |310\rangle_{ABC} - |060\rangle_{ABC}, \nonumber  \\
        \end{align*}
    \begin{align*}
       	|\varphi_{32}\rangle &= |000\rangle_{ABC} - |111\rangle_{ABC}, \nonumber  \\
      	|\varphi_{33}\rangle &= |0\rangle_{A}|0-6\rangle_{B}|6\rangle_{C}, \nonumber  \\
        |\varphi_{34}\rangle &= |0\rangle_{A}|0-6\rangle_{B}|7\rangle_{C}, \nonumber  \\
        |\varphi_{35}\rangle &=\left (\sum_{j=0}^{4}|j\rangle\right)_{A}\left (\sum_{j=0}^{6}|j\rangle\right)_{B}\left (\sum_{j=0}^{8}|j\rangle\right)_{C}, \nonumber  		
	\end{align*}
	where  $\omega = e^{\pi\sqrt{-1}}$.

	\section{Conclusion}
	\label{sec:5}
	In the field of quantum communication, a set of orthogonal quantum states with minimal nonlocality can be used to encode information. Since a set of orthogonal quantum states with minimal nonlocality cannot be perfectly distinguished by LOCC, an adversary cannot obtain useful information if the different particles of the quantum states encoding the information are transmitted separately. On the other hand, a set of orthogonal quantum states with minimal nonlocality can be perfectly distinguished by LOCC after removing one special state from this set. When information is encoded on this locally distinguishable subset of states, information extraction becomes possible via by LOCC. This cuts down the demand for quantum communication and reduces economic costs. Therefore, minimal nonlocality is of extremely important research value for quantum information processing and quantum communication.
	
	Existing studies have only provided construction methods for minimally nonlocal sets in equal-dimensional quantum systems. In this paper we propose a method to construct a set of orthogonal quantum states with minimal nonlocality  in both bipartite and tripartite quantun systems with unequal local dimensions. Our work enriches the foundation of the local distinguishability of quantum states.

\begin{acknowledgments}
	\vspace{-10pt}
    This work is supported by Natural Science Foundation of Shandong Province of China (Grant No. ZR2023MF080 and ZR2026MS1078) and Beijing Natural Science Foundation (Grant No. 4252014).
\end{acknowledgments}

\end{document}